\documentclass{article}

\usepackage{arxiv}

\usepackage[utf8]{inputenc} 
\usepackage[T1]{fontenc}    
\usepackage{hyperref}       
\usepackage{url}            
\usepackage{booktabs}       
\usepackage{amsfonts}       
\usepackage{nicefrac}       
\usepackage{microtype}      
\usepackage{lipsum}		
\usepackage{graphicx}
\usepackage[numbers]{natbib}

\usepackage{doi}

\usepackage{amssymb}

\usepackage{lineno}
\usepackage[utf8]{inputenc}
\usepackage{color}
\modulolinenumbers[5]
\usepackage{amsmath}
\usepackage{booktabs}
\usepackage{comment}
\usepackage{algorithm2e,algorithmic}
\usepackage{pgfplotstable}
\usepackage{array,multirow}
\usepackage{cancel}
\usepackage{tikz,tikz-inet,pgf,pgfplots}
\usepgfplotslibrary{dateplot}
\pgfplotsset{compat=1.13}

\usepackage{caption}
\usepackage{graphicx}
\usepackage{graphicx,rotating}
\usepackage[all]{xy}
\usepackage{subfig}
\usepackage{indentfirst,float}
\usepackage[output-decimal-marker={,}]{siunitx}

\title{Wavelet analysis and energy-based measures for oil-food price relationship as a footprint of financialisation effect}

\author{ \href{https://orcid.org/0000-0000-0000-0000}{\includegraphics[scale=0.06]{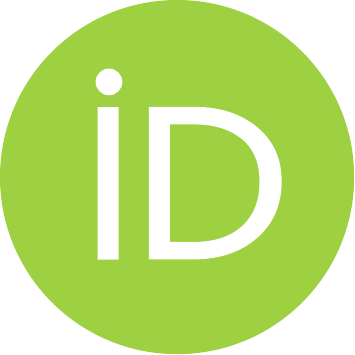}\hspace{1mm}Loretta Mastroeni}\thanks{} \\Dept. of Economics\\ Roma Tre University\\
Via Silvio D'Amico 77, 00145 Rome, Italy\\
              \texttt{loretta.mastroeni@uniroma3.it} \\
	\And
	\href{https://orcid.org/0000-0000-0000-0000}{\includegraphics[scale=0.06]{orcid.pdf}\hspace{1mm}Alessandro Mazzoccoli} \\
Dept. of Civil Engineering and Computer Science\\ University of Rome Tor Vergata\\ Via del Politecnico 1, 00133 Rome, Italy\\
              \texttt{alessandro.mazzoccoli@uniroma2.it}\\
			\And
	\href{https://orcid.org/0000-0000-0000-0000}{\includegraphics[scale=0.06]{orcid.pdf}\hspace{1mm}Greta Quaresima} \\
	Dep. of Methods and Models for Economics, Territory and Finance\\ University of Rome La Sapienza\\ Via del Castro Laurenziano 9, 00161 Roma RM\\
              \texttt{greta.quaresima@uniroma1.it}\\
			\And
	\href{https://orcid.org/0000-0000-0000-0000}{\includegraphics[scale=0.06]{orcid.pdf}\hspace{1mm}Pierluigi Vellucci} \\
	Dept. of Economics\\ Roma Tre University\\ Via Silvio D'Amico 77, 00145 Rome, Italy.\\
              \texttt{pierluigi.vellucci@uniroma3.it} \\
}

\renewcommand{\shorttitle}{\textit{arXiv} Template}

\hypersetup{
pdftitle={Wavelet analysis and energy-based measures for oil-food price relationship as a footprint of financialisation effect},
pdfsubject={q-bio.NC, q-bio.QM},
pdfauthor={David S.~Hippocampus, Elias D.~Striatum},
pdfkeywords={First keyword, Second keyword, More},
}

\begin{document}
\maketitle

\begin{abstract} 

In this paper we exploit the wavelet analysis approach to investigate oil-food price correlation and its determinants in the domains of time and frequency.
Wavelet analysis is able to differentiate high frequency from low frequency movements which correspond, respectively, to short and long run dynamics. We show that the significant local correlation between food and oil is only apparent and this is mainly due both to the activity of commodity index investments and, to a lesser extent, to a growing demand from emerging economies.
Moreover, the activity of commodity index investments gives evidence of the overall financialisation process. In addition, we employ wavelet entropy to assess the predictability of the time series under consideration at different frequencies.  We find that some variables share a similar predictability structure with food and oil.
These variables are the ones that move the most along with oil and food. We also introduce a novel measure, the Cross Wavelet Energy Entropy Measure (CWEEM), based on wavelet transformation and information entropy, with the aim of quantifying the intrinsic predictability of food and oil given demand from emerging economies, commodity index investments, financial stress, and global economic activity. The results show that these dynamics are best predicted by global economic activity at all frequencies and by demand from emerging economies and commodity index investments at high frequencies only.
\end{abstract}

\keywords{Wavelet analysis, Predictability, Wavelet entropy,  Oil-food co-movements, Financialisation}


\section{Introduction and literature review}


Crude oil is one of the most important energy commodities for its role in the global economy \citep{MASTROENI2020105036,ji2021intra}. Several developing countries depend on both crude oil and food imports, so co-movements between oil and food prices are a matter of great interest.

In the recent years, several explanations have been attributed to the relationship between oil and food prices.
As far as agricultural commodities are concerned, one explanation is that the oil and agriculture sectors are closely linked, since agricultural production is energy intensive.
Therefore, oil price increases result in higher costs for fertilisers, chemicals and shipping which in turn increase agricultural commodity prices \citep{articleDebdatta}. Another explanation is related to the production of biofuel that started in 2006 \citep{BASILI20121}. 


According to this theory, as oil prices rise, alternative energy sources are being developed,  such as bioethanol and biodiesel from corn and soybean, respectively. As a result, the increase in demand for these agricultural commodities drives up their prices \citep{WANG201422,articleDebdatta}.

 
A third explanation  is linked to  the growing demand from emerging economies, such as China and India. This demand has affected  a wide range of commodities from the energy and metals sectors and  may explain the co-movement between oil and food prices
\citep{WANG201422,doi:10.2469/faj.v68.n6.5}.

 
Global economic activity could also play a role, as it may be reasonable to think that global economic expansions could increase demand for oil and food, with both prices rising.  The impact of global economic activity has been studied by \cite{DONG2019134} only in relation to oil prices finding a significant relationship between the two.



A fourth explanation is related to the financialisation of commodity markets. The term financialisation is commonly referred to the tendency of commodity prices to behave like financial assets \citep{articleDebdatta,ADAMS2020104769}. The correlation between the price of financial assets and the price of commodities began to appear since the 2000s.
Commodity markets, in fact, before this period were segmented and not linked to the financial markets \citep{doi:10.2469/faj.v68.n6.5}. 


%

However, the discovery of a negative correlation between commodity returns and stock returns by \cite{doi:10.2469/faj.v62.n2.4083} led institutional  investors to consider commodities as a new asset class for diversification purposes
\citep{doi:10.2469/faj.v68.n6.5,ADAMS201593}.



Institutional investors use  commodity futures as a part of a broader portfolio strategy and often changed their investment positions from one commodity to another. This may induce co-movements between unrelated commodities as well as between oil and food prices 
\citep{ADAMS201593}.

However, most of the authors focus on finding evidence of a relationship between oil and food prices \citep{PAL20181032,REBOREDO2012456,articleDebdatta}, while only a few of them focus on its cause
\citep{ESMAEILI20111022,NAZLIOGLU2011488}. Along this line,  \cite{articleDebdatta} analysed the relation between Brent crude oil spot price and the food price index using wavelet analysis, finding a significant local correlation.\cite{CHENG2019422} investigated such relation using both threshold vector autoregression (TVAR) and threshold vector error correction (TVECM) models. They used monthly data of crude oil price index (an average of WTI, Brent and Dubai spot price) and food price index (a weighted average of the prices indices of meat, dairy products, grains, vegetables oils and sugar) spanning over 1990 to 2007. Their findings showed that co-movement of crude oil and food price had varying structural break features depending on the underlying regime. Following this thread, \cite{PAL20181032} analysed the impact of the food crisis on the interdependence between crude oil and food prices. They used detrended cross correlation analysis (DCCA) and considered monthly data on the food price index and Brent crude oil price ranging from January 1990 to July 2016. Their results suggested a strong interdependence between Brent crude oil and food price index. \textcolor{black}{ 
In \cite{bilgili2020estimation} and \cite{PEERSMAN2021103540} the authors also analysed the link between biofuel production and the oil price relationship.}
\textcolor{black}{In particular,\cite{bilgili2020estimation} investigated the impact of biofuel production on  the food prices in United States  using wavelet analysis and concluded  that there is  a significant relationship between the two in the short and long term. According to them, this finding can be useful to the government in designing sustainable energy and food policies.
 On the other hand, using a VAR framework, \cite{PEERSMAN2021103540} found that a fall in commodity supply increased oil prices.  However, their analysis suggested that this is not due to the biofuel industry but to information  discovery in financialized commodity markets and information frictions in global trade.}
 
 Besides, there are also dissenting voices on the co-movement between crude oil and food prices.
For example, \cite{REBOREDO2012456} found a weak dependence among Brent crude oil, wheat, corn and soybeans using copula functions. 
 With respect to the impact of financialisation, some authors as \cite{doi:10.1080/02692171.2015.1016406} and  \cite{ADAMS201593} did not properly analyse the impact of financialisation on the oil-food price co-movement, but analysed the correlation between commodity prices and stock prices. A more evident contribution on financialisation has been outlined in \cite{doi:10.2469/faj.v68.n6.5} who found evidence, using regression analysis, that  non-energy commodities in the two popular commodity indices, i.e. S\&P GSCI and DJ-UBSCI, were significantly correlated with oil prices compared to off-index commodities.
 
 \textcolor{black}{More recent papers, as for example the work of \cite{MOKNI2020101874}, analysed the ``causality''} \textcolor{black} {link between oil and food prices under different market conditions by detecting symmetrical and asymmetrical bi-directional causal relationships using the Granger causality test and causality in quantile test. }
 \textcolor{black}{As for the data, they used the world food price index and the average of the prices of the three benchmarks, namely WTI, Brent and Dubai. 
Their results showed that there is a causal impact  from oil price  to food price  under all market conditions, whereas the  causal impact of  food price on the oil price can be found only under extreme market conditions.}

\textcolor{black}{The effect of geopolitical risk is also important when studying the relationship between energy and agricultural markets.}
\textcolor{black}{To this purpose, the work  of \cite{TIWARI2021119584} used a copula-based approach on daily futures prices for oil and agricultural commodities traded on the New York Mercantile Exchange (NYMEX) in the period from  1990 to  2019. Their results} \textcolor{black}{suggest that the geopolitical risk has influenced the dependency between oil and agricultural commodities and, since geopolitical risk has negatively affected the oil market, agricultural commodities can hedge oil market risks.}
 
 \textcolor{black}{Most recent works on the oil-food price relationship take into consideration the role of the COVID-19 pandemic. In this strand of research, \cite{HUNG2021102236} analysed the relationship between oil and commodity markets using wavelet coherence and the spillover index approach. \textcolor{black}{They used daily data} of  WTI crude oil price and six agricultural grain commodities collected in the pre-COVID 19 period  (from February 2018 to January 2020) and in the post-COVID-19 period (from January 2020 to  May 2020). Their findings} \textcolor{black}{show that the intensity of  spillovers  from the crude oil market to the agricultural market is greater during the COVID-19 outbreak, shedding light on the evidence of increased spillovers during  periods of turmoil. }
\textcolor{black}{Their results also suggested that  the co-movement between these two markets is lower in} \textcolor{black}{the post-COVID-19 outbreak than before pandemic spread.}

\textcolor{black}{Also the work of \cite{UMAR2021102164} analysed the impact of COVID-19 on the  volatility of  commodity prices by using wavelet analysis. They used the Coronavirus worldwide pandemic index (PI) and commodity prices during the first seven months of 2020. Their } \textcolor{black}{results are important for hedging purposes, as the different degrees of coherence suggest a different degree of  correlation in the time and frequency domain  between  the Covid-19 pandemic and commodity market volatility.}

In Table \ref{table:summary_methods} we outline the methodological approach used both in the most recent documents and in those more related to our work.

In our paper, unlike most of the literature, we analyze the main causes of co-movement between oil and food prices using wavelet analysis tools.

More in details, we try  to answer to the following research questions:
\begin{enumerate}
    \item Is the oil-food price co-movement caused by the process of financialisation of the commodity markets?
    \item May the periods of increasing financial distress have played a role in the oil-food price co-movement?
    \item Is the oil-food price co-movement due to the increasing demand from emerging economies?
\end{enumerate}




To answer to these three research questions, we employ four tools from wavelet analysis: the wavelet coherence, the wavelet phase difference, the partial wavelet coherence and the partial phase difference.

From a methodological point of view, wavelet analysis has been extensively used ever since the pioneering works of \cite{haar1909theorie,Morlet1982I,Morlet1982II,Mallat1989IEEE,mallat1999wavelet} introduced, and then rigorously defined, the concept of wavelet. Examples of wavelet approach can be found in some field of engineering and applied mathematics, such as image processing problems, data engineering and denoising of various kind of signals \citep{BRUNI2020112467,chen2019coherent,Chen2020,BRUNI202073}, just to mention the most recent ones, but also in financial time series analysis \citep{BHUIYAN2021101971,SHEHZAD2021102163,MONGE2021102040,TIWARI2019118,MISHRA2019292,PAUL2019378}.

More in details, wavelet transforms are able to provide a time-scale representation of a time series. The scale allows to retrieve the information on frequency. Therefore, wavelet transforms are able to identify high frequency movements and low frequency movements in a time series. Low frequency movements capture the global information present in a time series, whereas high frequency movements capture transient information.

%

Wavelet analysis is important since  it  considers also the frequency domain. 
In fact, time domain analysis alone is not adequate to understand all data structures, especially for time series that include high-frequency or low-frequency movements.
\cite{Masset2015}.

In this case, wavelet coherence is  useful to have an inspection of the local correlation in time and scale between two time series where the scale is linked to the frequency. Moreover, wavelet phase difference is useful to analyse the sign of the local correlation. Partial wavelet coherence, on the other hand, is used to see if the correlation between two time series is due to the effect of another series.
This is obtained by controlling the wavelet coherence between two series with respect to a third one.

 To answer the first research question, as a first step we employ the wavelet coherence between the food price index  and the oil price index. After that, we employ the partial wavelet coherence between the same time series and control for the effect of the S\&P GSCI commodity index. If most of the local correlation disappears once we check this series, then we have evidence that  most of the correlation is due to the S\&P GSCI commodity index. 
Therefore, the oil-food price correlation is simply due to commodity index investments, thus providing evidence of the impact of financialisation on the oil-food price relationship.  




With respect to the second research question, we employ the wavelet coherence between the food price index and the oil price index and then we employ the partial wavelet coherence controlling for TED spread. The TED spread is used to assess the contribution of systemic financial distress on the oil-food price relationship. Therefore, if most of the local correlation between food-oil disappears when we control for the TED spread, we have evidence that the oil-food price correlation is simply due to the systemic financial distress.

For the third research question, we employ the wavelet coherence between the food price index and the oil price index and then we employ the partial wavelet coherence between the same series by  checking the effect of the MSCI Emerging Markets Index (MSCI). This is commonly used as a proxy for the growth of emerging economies \citep{doi:10.2469/faj.v68.n6.5}. Therefore, if most of the local correlation between food and oil disappears when we check the growth of emerging economies, then we have evidence that MSCI is the factor that mostly contributes to the oil-food price relationship.  Besides, we used the Kilian economic index to assess if oil-food price correlation can be explained by global economic activity rather than just by demand 
from emerging economies. For this purpose, we adopted the same methods described above.

Moreover, we investigate how predictable the time series under analysis are and whether  this predictability structure is present for variables that  mostly co-move  with both food and oil. 
To this end, the concept of entropy is combined with that of wavelets that leads to the definition of the wavelet entropy. We recall that lower (higher) entropy values are associated to predictable (unpredictable) time series. Since one of the advantages of wavelet transform is a time-scale representation of the original time series, wavelet entropy gives us a measure of the predictability as the scale changes. Wavelet entropy is fundamental to define the concept of Wavelet Entropy Energy Measure (WEEM), which is used to assess the intrinsic predictability of time series at different scales. We adopted a slightly different approach to the definition of WEEM than that of \cite{MarwanPerc2020}. \cite{MarwanPerc2020} need a time series representing white noise, while in our paper we need only its entropy (which is maximal if all the outcomes are equally likely). Moreover, in our paper, given two time series, we take the methodological approach a step further, with the goal of quantifying the inherent predictability of one given the other and vice versa.
To this aim we introduce here the Cross Wavelet Energy Entropy Measure (CWEEM) generalizing the WEEM proposed by \cite{MarwanPerc2020}.


The results of our analysis show that the activity of commodity index investments is primarily responsible for the apparent co-movement between food and oil. However, demand from emerging economies also plays an important role, but to a lower extent. Furthermore, commodity index investments and demand from emerging economies are positively correlated both with food and oil. More in details, the positive correlation is long-term, so it refers to global information on the dynamics of time series.
Conversely, the co-movement among TED spread, oil and food is short term, thus it is associated to a transient information on the dynamics of time series.
TED spread is only responsible for co-movement of oil and food prices during the financial crisis, so its effect cannot be considered as one of the leading causes.
Besides,the results of WEEM shows that both oil and food  share a similar  predictability structure with GSCI and MSCI. Actually, GSCI and MSCI are the variables that mostly co-move with oil and food.  Conversely, the results of CWEEM show that MSCI e GSCI helps to predict high frequency movements in food and oil price series, whereas global economic activity predicts both high frequency and low frequency movements.

Our contribution to the literature is outlined in the following points:

\begin{itemize}
    \item  Differently from the work by \cite{articleDebdatta},  we found that the oil-food price co-movement is only apparent. Indeed, most of the co-movements are due to the activity of commodity index investments;
    \item  We have shown the impact of demand from emerging economies on the oil-food price co-movement.  It should be noted that demand from emerging economies has contributed less to the co-movement of oil and food prices than investment activity in commodity indices;
    \item We found that the TED spread is not the main responsible for the oil-food price co-movement; 
    \item Compared to the work of \cite{DONG2019134}, we study  the impact of global economic activity not only on oil but also on food;
   
\end{itemize}



We also provide some policy implications on the basis of our results. 



The remainder of this article is structured as follows. 
Section \ref{data} describes the data used in the current study and discuss the  corresponding summary  statistics. Section \ref{sec:method} is devoted to the description of the tools we employ in our analysis, while Section \ref{results} discusses the results. In addition, Section \ref{conclusions} provides conclusions and some policy implications on the basis of our results. Finally, in the appendix we give a deeper description of the algorithms employed for the WEEM and CWEEM computation.

\begin{table}[p]
\centering
\resizebox{\textwidth}{!}{%
\begin{tabular}{|p{2.5cm}|p{5cm}|p{3cm}|p{4cm}|}\hline
{Authors} & {Data} & {Methods} & {Main findings} \\\hline

{\cite{PAL20181032}} & {Monthly  spot price of Brent and food price index ranging from 1990 to 2016} & {DCCA} & {Strong interdependence between Brent and food price index} \\
{\cite{REBOREDO2012456}} & {Monthly spot prices of  Brent, wheat, corn and soybean} & {Copula functions} & {Weak dependence between crude oil and food prices} \\
{\cite{articleDebdatta}} & {Monthly spot price of  Brent and food price index over the period 1990-2016} & {Wavelet analysis} & {Brent crude oil spot price leads the food price index in the period 2006-2008}  \\
\cite{doi:10.1080/02692171.2015.1016406}& {Agricultural commodity prices,S\&P 500 index, Kilian index, TED spread} & {Time varying correlations and Granger causality test} & {Combination of financialisation and finacial turmoil increase the correlation between commodity and stock prices} \\
{\cite{ADAMS201593}} & {Daily data for S\&P 500 index, MCSI world index, GSCI excess return values for corn, wheat, crude oil, livestock, heating oil, alluminium, copper, gold over the period 1995 - 2013} & {Change point correlation and SDSVar} & {Investment styles are responsible for the correlation between commodity and stock market prices}\\
\cite{CHENG2019422} & {Oil price index and food price index over the period 1990-2007} & {TVAR,TVECM} & {Evidence of oil-food co-movement}\\
{\cite{DONG2019134}} & {Monthly spot price of WTI and Brent and the Kilian index over the period 1985-2018} & {Wavelet analysis} & {Relationship between oil prices and Kilian index is stronger at higher frequencies}\\
\bottomrule
\end{tabular}}
\caption{Broad classification of most recent related literature concerning oil-food price relationship}
\label{table:summary_methods}
\end{table}


\section{Data description and preliminary statistics}
\label{data}

In this section, we describe the data used in the current study and discuss the corresponding summary statistics. We choose a monthly dataset that spans from June 2000 to February 2020 for a total of 237 observations. The decision to start the dataset from 2000 is due to the current availability of some data from this date on-wards.



We used monthly observations of the oil price index and the food price index. The oil price index  can be thought as the simple average of the spot prices of Brent, WTI and Dubai \citep{CHENG2019422}. On the other hand, the food price index includes the mean of the price indices of diary, cereals, vegetable oil, meat and sugar \citep{articleDebdatta}.  The data on the oil price index and the food price index are obtained from the Federal Reserve Bank of Saint Louis\footnote{www.stlouisfed.org}.


We used also monthly observations of the the Goldman Sachs Commodity (GSCI) total return index. The GSCI index reports futures prices for a basket of commodities belonging to energy, agricultural and metal sector \citep{doi:10.1146/annurev-financial-110613-034432}. The GSCI index is one of the largest indices by market share and to each commodity in the index is assigned a weight. 
In addition, the logic of weight assignment in the GSCI index is based on the worldwide production of each commodity
\citep{doi:10.2469/faj.v68.n6.5}. To the energy sector, and oil in particular,  is given the highest weight. 
We used also monthly observations of the MSCI Emerging Market Index to approximate the growth of emerging economies. The data on the GSCI total return index and MSCI Emerging Market Index are downloaded from Thomson Reuters Datastream. 

Furthermore, we employed monthly observations of the TED spread. The data on the TED spread are obtained from the Federal Reserve Bank of Saint Louis. The TED spread is given by the difference between the 3-Month London Interbank Offered Rate (LIBOR) and the 3-Month Treasury Bill secondary market rate and is commonly used as a measure of systemic  financial distress \citep{doi:10.1080/02692171.2015.1016406}.
We also used the Kilian economic index which acts as a proxy for  global economic activity. This index is based on representative dry cargo single-voyage ocean freight rates and there are different dry cargoes that include grain, oil seeds, coal, iron ore, fertilizer and scrap metal \citep{DONG2019134}. 
The Kilian economic index is downloaded from the following website \footnote{https://sites.google.com/site/lkilian2019/research/data-sets}.

Table \ref{table:statistics} shows  the descriptive statistics of our data. The time series are all positive skewed except for MSCI index and food price index. Furthermore, the series are all platykurtic except for the Ted spread. The Jarque-Bera statistic rejects the null hypothesis that the data are normally distributed. The Augmented Dickey Fuller (ADF) test does not provide sufficient evidences to reject the null hypothesis of a unit root in the series. Besides, the Ljung-box statistic shows that there is temporal dependence in our series up to lag $1$. 


\begin{table}[hbt!]
\centering
\resizebox{\textwidth}{!}{%
\begin{tabular}{lccccccccc}
\toprule
&{Mean} & {Std. Dev.} & {Min} & {Max}& {Skewness} & {Kurtosis} & {Jarque-Bera} & {Ljung-Box}& {ADF}  \\
\midrule
{Oil} & {139.22} & {55.44} & {47.55} & {264.61} & {0.25} & {-1} & {11.94***} & {229.13***} & {-1.75} \\
{Food} & {94.56} & {20.77} & {56.21} & {132.80} & {-0.21} & {-1} & {11.30***} & {232.63***} & {-1.72} \\
{GSCI} & {320.48} & {104.32} & {141.14} & {612.92} & {0.15} & {-0.69} & {5.35*} & {229.94***} & {-1.72}\\
{MSCI} & {811.19} & {291.65} & {250.29} & {1327.89} & {-0.57} & {-1.06} & {23.97***} & {230.49***} & {-2.50}\\
{TED} & {0.42} & {0.39} & {0.11} & {3.31} & {3.59} & {17.35} & {3548.80***} & {167.39***} & {-2.73} \\
{Kilian} & {11.34} & {73.07} & {-159.64} & {190.73} & {0.45} & {-0.44} & {9.89***} & {217.55***} & {-2.51} \\
\bottomrule
\end{tabular}}
\caption{Descriptive statistics. In  Table \ref{table:statistics} ***, ** and * denote rejections of null hypothesis at  1 \%,5\% and 10 \% significance levels , respectively.}
\label{table:statistics}
\end{table}

\section{Methods}
\label{sec:method}

Fourier analysis is one of the commonly used methods to analyse periodicity in the frequency domain. In particular,
Fourier transform (FT) uses sine and cosine functions to reconstruct a signal or a time series and gives information about their global frequency distribution. Since time information is not considered, FT is most suitable for time series that are generated by time-invariant systems (\cite{merry2005wavelet}). 
The loss of time information makes it difficult to identify both transient relations and structural changes and discontinuities in the series under study \citep{doi:10.1111/joes.12012,Bloomfield_2004,merry2005wavelet}. This drawback is overcome by wavelet analysis, since, as we will see later in this section, the continuous wavelet transform (CWT) preserves both time and frequency information by decomposing  the original time series into a  wavelet function parameterised in terms of time location and scale \citep{Bloomfield_2004,doi:10.1111/joes.12012}. 
Then, it is usual to refer to this as a time-frequency analysis.

A wavelet transform is then a linear transformation in which the basis functions are scaled and shifted versions of one function, called the mother wavelet. The scale allows to retrieve frequency information through a relationship that depends on the type of mother wavelet used. The scale is used to control the length of the CWT and it is varied endogenously: it is stretched (compressed)  to capture low (high) frequency movements in the time series
\citep{doi:10.1111/joes.12012}.
Low frequency movements provide information on the global frequency of the time series, whereas high frequency movements reflect transient information. Since global information is given by the frequency that contributed most to the dynamics of the time series, it can be interpreted as a long-run component, whereas high-frequency movements are short-lived and can be associated with short-run dynamics.
Despite Fourier analysis assumes that the underlying process evolves as a stationary process over time, wavelet transform can work on both non-stationary and locally stationary series \citep{ROUEFF2011813,grinsted:hal-00302394,10.1175/1520-0477(1995)076<2391:CSDUWT>2.0.CO;2}).


\textcolor{black}{In particular,  wavelet  has the peculiarity and power to localise stationary and non-stationary signals of a time series in both time and frequency domains at the same time.
In fact, following high or low frequency movements, it can grasp all possible co-movements between the variables of a series at different times, considering possible different frequencies and different periods  (\cite{bilgili2019revisited},\cite{bilgili2020estimation}).}

\textcolor{black}{Furthermore, wavelet analysis differs from panel data models and VAR models for two main reasons: wavelet analysis is a non-parametric method and works in both the time and frequency domains, whereas panel data models are used for  data which  involve the observations of a number of different variables, each over a range of time periods  and VAR models work only in the time domain.
Moreover wavelets are often used for time series data but there are some studies as the works by \cite{SALDIVIA2020115207} and \cite{KRISTJANPOLLERR2018640} that use wavelets for panel data as well.} \textcolor{black}{In their works, they analyse the relationship between gross domestic product (GDP) and energy consumption, using wavelet transformation to analyse data co-movements  and causality, employing data panel, since they provide greater accuracy in statistical inference, and considering cross-sectional dependence of the data.}

Intuitively, we could think of the wavelet transform as a tool that allows us to decompose the time series we want to analyze into several temporal segments and to carry out the frequency analysis of that section of the series. 
Besides, wavelet transform can work also on non stationary series, since it breaks up the time series in segments where frequencies are approximately constant and carries out independent analysis on that portion. 
However, time-frequency analysis is affected by the Heisenberg uncertainty principle and this applies  \textcolor{black}{ particularly for the wavelet transform: as resolution increases over time, frequency resolution is lost. 
Actually, time and frequency accuracy are hyperparameters of the wavelet tool chosen according to the time-frequency dynamics of the analyzed time series. } 

Some additional tools can be derived from wavelet transforms in order to work on more than one time series. These tools are the cross wavelet transforms, the wavelet coherence and the wavelet phase difference, which are used to quantify the local covariance, the  local correlation and the lead/lag relationship among two time series, respectively. 

Wavelets can also be combined with entropy to quantify the determinism component present in a time series and which determines its predictability. To this purpose, discrete wavelet transforms (DWT) are used. It can be shown that DWT can be obtained from the continuous wavelet transform when we restrict the scale $s$ to some discrete values $2^j$. 

An important advantage of wavelet analysis over traditional time series approaches concerns the fact that it is able to consider relationship among variables in both time and frequency domain. In addition,  traditional time series approaches, working only on time domain, divide time scale into short and long run (e.g. cointegration analysis, vector autoregressive and vector error correction model), whereas wavelet analysis is able to discover multiscale relationship between variables \citep{DONG2019134}.

\subsection{Continuous wavelet transform}
Let us consider a time series $x(t)$  and a wavelet function $\psi(t)$. The continuous wavelet transform (CWT) $W(\tau,s)$ at scale $s>0$ and translation parameter $\tau\in {\mathbb  {R}}$  maps the original time series into a dilated and translated version of the mother wavelet. In particular, it is defined as follows:
\begin{equation}
\label{eq:cwt}
 W(\tau,s)=\frac {1}{\sqrt{s}}\int _{-\infty }^{\infty }x(t){\overline {\psi }}\left({\frac {t-\tau}{s}}\right)\,dt, 
\end{equation}
where $\psi(t)$, called the mother wavelet, is a continuous function both in time and  frequency domain and where the symbol overline represents the operation of complex conjugate used in the case of complex wavelet. As for the properties that the mother wavelet must satisfy see e.g. \cite{lee1994wavelet,VACHA2012241,merry2005wavelet,10.1175/1520-0477(1995)076<2391:CSDUWT>2.0.CO;2,DONG2019134,aguiar2011oil}. 


We can derive from  Eq. (\ref{eq:cwt}) a measure of the localized variance of the time series, that is the so called wavelet power spectrum (WPS) which is defined as follows: 
\begin{equation}
     WPS_x=|{W_{x}(\tau,s)}|^2.
\end{equation}

\subsection{Morlet wavelet}

There are several wavelet functions available to play the role of the mother wavelet. In particular, in this paper, we make use of the Morlet wavelet, that is defined as follows:
\begin{equation}
\psi(t)=\pi^{-\frac{1}{4}}e^{-\frac{1}{2}t^{2}}\, e^{i \omega_0 t}
\end{equation}
where $w_0$ is a dimensionless frequency and $t$ is a dimensionless time.

 The Morlet wavelet is interesting for three main reasons: it is a complex function, it provides a simple conversion of scales into frequency and it has an optimal joint time-frequency localization.

The fact that the wavelet is a complex function allows us to recover  information on amplitude and phase, both of which are fundamental to study time delay between the oscillations of two time series \citep{aguiar2011oil}.
The simple conversion of scale into frequency occurs when $w_0$=6. This value ensures  that the scale $s$ is inversely related to frequency $f$ ($f$ $\approx$ $1/s$) \citep{aguiar2011oil,doi:10.1111/joes.12012}.

The application of the Heisenberg uncertainty principle in the context of the wavelet analysis results in the fact  that it is not possible to achieve high accuracy of time and frequency information at the same time.
As a consequence, there is always a trade-off, and, since the Morlet wavelet has an optimal trade-off between accuracy of frequency and time information, it is often preferred with respect to other wavelet functions \citep{grinsted:hal-00302394,doi:10.1111/joes.12012}.

\subsection{Cross wavelet transform and wavelet coherence}

Let us consider \textcolor{black}{two input time series } $\{X_t\, :\, t = 0,\dots , N - 1\}$ and $\{Y_t\, :\, t = 0,\dots , N - 1\}$, with continuous wavelet transforms $W_x(\tau,s)$ and $W_y(\tau,s)$. The cross-wavelet transform of the two series $\{X_t\}$ and $\{Y_t\}$ is defined as follows:
\begin{equation}
\label{eq:cross}
     W_{x,y}(\tau,s)= W_x(\tau,s)\overline{W_y(\tau,s)},
\end{equation}
Starting from this equation, we can define the cross-wavelet power spectrum (CWS):
 \begin{equation}
    CWS_{x,y}=|{W_{x,y}(\tau,s)}|.
     \label{eq:crosswts}
\end{equation}
CWS can be seen as a measure of the localized covariance between two time series \citep{doi:10.1175/1520-0477(1998)079<0061:APGTWA>2.0.CO;2}. Let us remark that areas with high common power, as well as relative phase in time frequency space, are revealed by Eq. (\ref{eq:crosswts}). However, the cross wavelet power spectrum can show high common power even if the series are realisations of independent processes and may lead to spurious significance tests. 



Therefore, another measure as the wavelet coherence is often preferred. According to  \cite{doi:10.1111/joes.12012}, wavelet coherence of two time series is defined as follows
 \begin{equation}
     R_{x,y}(\tau,s)=\frac{  |S ({W _{x,y}(\tau,s)})|}{S(|W_x(\tau,s)|^{2})^{1/2} S(|W_y(\tau,s)|^{2})^{1/2}}.
     \label{eq:wavecoh}
\end{equation}
Eq. (\ref{eq:wavecoh}) is obtained from the cross-wavelet power spectrum normalized by the wavelet-power spectrum of the two series and $S$ represents a smoothing operator in both time and scale. 
\cite{10.1175/1520-0442(1999)012<2679:ICITEM>2.0.CO;2} and \cite{grinsted:hal-00302394} defined the smoothing operator as described in the following:
\begin{equation}
\label{eq:smooth}
    S(W(\tau,s))=S_{scale}(S_{time}(W(\tau,s))),
\end{equation} where
\begin{subequations}
\begin{equation}
     S_{time}(W(\tau,s))|_s=\left( {W(\tau,s)}*c_1^{-\frac{t^2}{2s^2}}\right)\Biggl|_s,
\end{equation}
\label{timesmo}
\begin{equation}
     S_{scale}(W(\tau,s))|_\tau=\left( {W(\tau,s)}*c_2\Pi(0.6s)\right)|_\tau,
\end{equation}
\label{sclalesmo}
\end{subequations}
$c_1$ and $c_2$ are normalisation constants and $\Pi$ is the rectangular function. Moreover, in Eq. (\ref{timesmo}a) and in Eq. (\ref{sclalesmo}b), the symbol * denotes the convolution operator   \citep{grinsted:hal-00302394,helmanexploring}.

The time smoothing used in Eq. (\ref{timesmo}a) employs a filter identified by  the absolute value of the wavelet function at each scale and normalized to have a total weight of unity \citep{pacchetto}. For the Morlet wavelet, this is represented by a Gaussian function. The scale smoothing in Eq. (\ref{sclalesmo}b) is employed using a Boxcar function of width $0.6$. The width corresponds to the decorrelation scale of the Morlet wavelet \citep{pacchetto}.  
As suggested by \cite{10.1175/1520-0442(1999)012<2679:ICITEM>2.0.CO;2}, we apply the smoothing in both time and scale. 

The value of the  wavelet coherence falls in the interval $[0,1]$ and is a measure of the local  correlation between two time series at location $\tau$ and scale $s$. A strong linear relationship is found when the wavelet coherence is close to one and indicates a strong co-movement,  whereas non-linear relationship is present when the value of the wavelet coherence is zero \citep{BASTA2018204}. 

It is important to note that wavelet coherence does not distinguish between positive or negative co-movements, therefore the relative phase (or phase difference) is needed since it provides information about the delay between two time series at location $\tau$ and scale $s$ \citep{DONG2019134,maraun:hal-00302384}.   



\subsection{Partial coherence and partial phase difference}

Partial coherence and partial phase difference  are useful to investigate the localized correlation along with the lead-lag relationship among two series $\{X_t\}$ and $\{Y_t\}$ by controlling for the effect of another series $\{Z_t\}$ \citep{DONG2019134}. In particular, the partial wavelet coherence (PWC) $R_{x,y|z}(\tau,s)$ is defined as follows
\begin{equation}
     R_{x,y|z}(\tau,s)=\frac{  |R_{x,y}(\tau,s)-R_{x,z}(\tau,s)\overline{R_{y,z}(\tau,s)}|}{\Biggl(1-(R_{x,y}(\tau,s))^{2}\Biggr)^{1/2} \Biggl(1-(R_{y,z}(\tau,s))^2\Biggr)^{1/2}},
     \label{eq:partialcoh}
\end{equation}
where $R_{i,j}(\tau, s)$ is defined according to Eq. (\ref{eq:wavecoh}). Instead, the partial phase difference is defined as follows
\begin{equation}
   \theta_{x,y|z}= \left(\frac{\mathfrak{I}{(C _{x,y|z}}(\tau,s))}{\mathfrak{R}{(C _{x,y|z}}(\tau,s))}\right),
\label{eq:partialphase}
\end{equation}
where  $C_{x,y|z}$ is the complex number whose absolute value is equal to $R_{x,y|z}$ \citep{DONG2019134,doi:10.1111/joes.12012}.

\subsection{Wavelet phase difference}
\label{phase}

In order to detect the sign of co-movements we need to introduce the wavelet phase difference. The latter is derived using the real and the imaginary part of cross wavelet transform defined by Eq. (\ref{eq:cross}) in the Appendix \citep{Bloomfield_2004}. More in details, the wavelet phase difference is defined as follows
\begin{equation}
\label{eq:phase} 
\theta_{x,y}= \arctan\left(\frac{\mathfrak{I}(S ({W _{x,y}(\tau,s))}}{\mathfrak{R}(S({W _{x,y}(\tau,s))}}\right)
\end{equation}
where $\theta_{x,y}$ $\in$ $[-\pi, \pi]$, and with $\mathfrak{R}[z]$ and $\mathfrak{I}[z]$ we refer to the real and the imaginary part of $z\in \mathbb{C}$. According to the value of  $\theta_{x,y}$, it is possible to establish if the series co-move positively (negatively) and which series is leading (lagging) the other one. More specifically, several cases can be distinguished:
\begin{itemize}
    \item if $\theta_{x,y}$= $0$, the two series are in-phase (positive co-movement)  and no lead/lag relationship is present. The arrow will be pointing to the right ($\longrightarrow$); 
    \item if $\theta_{x,y}$ $\in$ $(0, \frac{\pi}{2})$, the two series are in-phase (positive co-movement) with $\{X_t\}$ leading $\{Y_t\}$. The arrow will be pointing up and right ($\nearrow$);
    \item if $\theta_{x,y}$ $\in$ $(\frac{\pi}{2}, \pi)$, the two series are out of phase (negative co-movement) with $\{Y_t\}$ leading $\{X_t\}$. The arrow will be pointing up and left ($\nwarrow$);
    \item if $\theta_{x,y}$ $\in$ $(-\frac{\pi}{2}, 0)$, the two series are in-phase (positive co-movement) with with $\{Y_t\}$ leading $\{X_t\}$. The arrow will be pointing  down and right ($\searrow$);
    \item if $\theta_{x,y}$ $\in$ $( -\pi,-\frac{\pi}{2})$, the two series are out of phase (negative co-movement) with  $\{X_t\}$ leading $\{Y_t\}$. The arrow will be pointing down and left ($\swarrow$).
\end{itemize}


 \subsection{Discrete Wavelet Transform}
When we restrict the scale $s$ to some discrete values $2^j$ we get the discrete wavelet transform (DWT).
Let $X_0, X_1,\dots, X_{N-1}$ represent a time series of $N$ real-valued variables (henceforth we will denote such a series as either $\{X_t\, :\, t = 0,\dots , N - 1\}$ or just $\{X_t\}$ if it is clear what values the dummy index $t$ can assume). We also let $\textbf{X}$ represent the column vector of length $N$ whose nth element is $X_n$.
 
 The DWT of $\{X_t\}$ is an orthonormal transform. By $\{W_n \, :\,  n = 0,\dots, N - 1\}$ and $\textbf{W}$ we denote, respectively, the DWT coefficients and the column vector of length $N = 2^J$ whose nth element is the nth DWT coefficient $W_n$. Hence, $J$ is the largest DWT level for sample size $N=2^J$. We can write
\begin{equation}
    \textbf{W}=\mathcal W \, \textbf{X}
\end{equation}
where $\mathcal W$ is an $N \times N$ real-valued matrix, also called \emph{discrete wavelet transform matrix}, and satisfying $\mathcal W^T \mathcal W=I_N$ ($I_N$ is the $N\times N$ identity matrix). Similar to the orthonormal discrete Fourier transform, orthonormality implies that $\textbf{X}=\mathcal W^T \, \textbf{W}$ and $\|\textbf{W}\|^2=\|\textbf{X}\|^2$. Therefore, $W_n^2$ represents the contribution to the energy attributable to the DWT coefficient with index $n$, as described by \cite{percival2000wavelet}. As we will see in the following, this result will allow us to define the concept of \emph{wavelet entropy}.

 A precise definition of DWT is formulated as an algorithm that allows $\mathcal W$ to be factored in terms of very sparse matrices. This  algorithm is known as the pyramid algorithm and was introduced in the context of wavelets by \cite{Mallat1989IEEE}. It is based on the wavelet synthesis of $\textbf{X}$ indicated by the following equation:
\begin{equation}
    \textbf{X}=\mathcal W^T \, \textbf{W}=\sum_{j=1}^J \mathcal W_j^T \textbf{W}_j+ \mathcal{V}_J^T \textbf{V}_J
\end{equation}
where we define the $\mathcal W_j$ and $\mathcal V_J$ matrices by partitioning the rows of $\mathcal W$ commensurate with the partitioning of $\textbf{W}$ into $\textbf{W}_1,\dots,\textbf{W}_J$ and $\textbf{V}_J$. Thus the $\frac{N}{2^j} \times N$ matrix $\mathcal W_j$ is formed from the $n = \frac{j-1}{2^{j-1}}N$ up to $n = \frac{2j-1}{2^j}N - 1$ rows of $\mathcal{W}$ for $j=1,\dots,J-1$. The $1\times N$ matrices $\textbf{W}_J$ and $\textbf{V}_J$ are the last two rows of $\mathcal W$. We thus have 
\begin{equation}
\label{eq:Wdecomp}
    \mathcal W= \left[\begin{array}{c}
         \mathcal{W}_1\\
         \mathcal{W}_2\\
         \vdots\\
         \mathcal{W}_J\\
         \mathcal{V}_J
    \end{array} \right]\, .
\end{equation}
The first $N/2$ rows of this matrix are obtained through a DWT \emph{wavelet filter}, which is built upon an infinite sequence $\{h_l\}$ with at most $L$ nonzero values. A wavelet filter must satisfy the following three basic properties: 
\begin{align}
    &\sum_{l=0}^{L-1}h_l=0 \notag \\
    &\sum_{l=0}^{L-1}h_l^2=1 \notag \\
    &\sum_{l=0}^{L-1}h_l\, h_{l+2n}=0\, .
\end{align}
Let then $\{h_l\, : \, l=0,\dots,L-1\}$ be a wavelet filter of even width $L$. Through the wavelet filter, we construct the first $N/2$ rows of $\mathcal W$, i.e. the matrix $\mathcal{W}_1$ in the decomposition of $\mathcal{W}$ shown in Eq. (\ref{eq:Wdecomp}). In preparation for forming the last $N/2$ rows of $\mathcal W$ via the pyramid algorithm, we now define a second filter.

 The required second filter is the \emph{quadrature mirror} filter (QMF as described by \cite{percival2000wavelet}) $\{g_l\}$ that corresponds to $\{h_l\}$:
\begin{equation}
    g_l := (-1)^{l+1} \, h_{L-l-1}\quad \Leftrightarrow  \quad h_l := (-1)^{l} \, g_{L-l-1}\, .
\end{equation}
The filter $\{g_l\}$ is known as the \emph{scaling filter}. It can be shown that the scales associated with the outputs of the wavelet and scaling filters differ by a factor of two. Therefore we use $\lambda_j = 2^j$ to denote the scale of the output from the scaling filter, whereas we use $\tau_j = 2^{j-1}$ to denote the scale associated with the output from the wavelet filter \citep{percival2000wavelet}.

 Thanks to wavelet and scaling filters, it is possible to define the jth stage of the pyramid algorithm. Let $V_{0,t}=X_t$, the jth stage input is $\{V_{j-1,t}: t=0,\dots,N_{j-1}-1\}$, where $N_j=N/2^j$. This input is the scaling coefficients associated with averages over scale $\lambda_{j-1}=2^{j-1}$. The jth stage outputs are the jth level wavelet and scaling coefficients: 
\begin{equation}
\label{eq:jthoutput}
W_{j,t}=\sum_{l=0}^{L-1}h_l V_{j-1,2t+1-l} \, \mod N_{j-1}, \quad V_{j,t}=\sum_{l=0}^{L-1}g_l V_{j-1,2t+1-l} \, \mod N_{j-1}
\end{equation}
for $t = 0,\dots, N_j - 1$. The wavelet coefficients for scale $\tau_j=2^{j-1}$ are given by
\begin{equation*}
    \textbf{W}_j=\left[W_{N-N_{j-1}},W_{N-N_{j-1}+1},\dots,W_{N-N_{j}-1}\right]^T=\left[ W_{j,0}, W_{j,1},\dots, W_{j,N_j-1}\right]^T
\end{equation*}
We are now ready to describe the pseudo-code that can be used to compute the DWT using the pyramid algorithm. Denote the elements of $\textbf{W}_j$ and $\textbf{V}_J$ as $W_{j,t}$, $t=0,\dots,N/2^j-1$ and $V_{J,t}$, $t=0,\dots,N/2^J-1$. Given the vector $\textbf{V}_{j-1}$ of even length $M = N/2^{j-1}$, the Algorithm \ref{alg:DTW} --- contained in \cite[Comments and Extensions to Section 4.6]{percival2000wavelet} --- computes the vectors $\textbf{W}_{j}$ and $\textbf{V}_{j}$ each of length $M/2$ and it is based on Equation (\ref{eq:jthoutput}). In this way we obtain the component vectors of the DWT, i.e. $\textbf{W}_1,\dots,\textbf{W}_J$ and $\textbf{V}_J$.

\subsection{Wavelet energy entropy measure (WEEM)}
\label{WEEM}

In this paper we employ wavelet entropy to assess the predictability of the time series under consideration. Wavelet entropy is fundamental to define the concept of Wavelet Entropy Energy Measure (WEEM), which is used to assess the intrinsic predictability of the time series at different scales. In order to introduce WEEM, we have to start from the concept of energy for orthonormal basis.

The concept of energy for an orthonormal basis is linked with well-established notions derived from the Fourier theory. In particular, an orthonormal wavelet basis converges in norm, and the energy of a function (signal) is defined as the integral of its absolute value \citep{Loreti2018}. Then, we can define the relative energy of the wavelet coefficients at each scale $j$ as follows
\begin{equation}
\label{eq:en_rel}
E_j=\frac{\|\textbf{W}_j\|^2}{\sum_{j=1}^J \|\textbf{W}_j\|^2}\, ,
\end{equation}
(see \cite{alcaraz2012application}), which defines by scales the probability distribution of the energy. Clearly $\sum_{j=1}^J E_j=1$ and the distribution $\{E_j\}$ can be considered as a time-scale density. This tool has been employed for detecting and characterizing specific phenomena in time and frequency domains, as shown in \citep{alcaraz2012application,ROSSO200165,BRUNI20122891,BRUNI202096}.

 Starting from the previous definition, we can introduce the   
Wavelet Entropy (WE) as
\begin{equation}
\label{eq:WE}
WE=-\sum_{j=1}^J E_j\, \ln\left(E_j\right)\, .
\end{equation}

The  Wavelet Entropy is based on the concept of Shannon Entropy introduced by \cite{Shannon1948} --- ``the father of information theory'' --- and widely used in sciences including energy finance/economics \citep{DIMPFL2018368,Benedetto2019,BENEDETTO2016302}. It provides useful information about the underlying dynamical process of the time series, and, in particular, it is a measure of the degree of order/disorder of the time series. In fact, periodic mono-frequency time series, with a narrow band spectrum, are also the most ordered and regular and a wavelet representation of these kind of time series is characterized by one unique wavelet resolution level which includes the representative frequency. For this special level the relative wavelet energy will be almost one, whereas all the other relative wavelet energies will be almost zero. Consequently, the total WE will be close to zero or very low.
A totally random process, instead, generates a time series whose wavelet representation has significant contributions from all frequency bands. 
Moreover, one could expect all contributions to be of the same order and the WE to reach their maximum values \citep{ROSSO200165}.


Recently, \cite{MarwanPerc2020} have introduced a novel measure, the Wavelet Entropy Energy Measure (WEEM), based on wavelet transformation and information entropy for the quantification of \emph{intrinsic predictability} of time series. 
According to \cite{Pennekamp2019} predictability can be classified into two different types:
\begin{itemize}
    \item realized predictability, which is the achieved predictability of a system from a given forecasting model (in other words, it indicates forecast performance of models);
    \item intrinsic predictability, which is the maximum achievable predictability of a system, see also \cite{Beckage2011}.
\end{itemize}
To quantify the intrinsic predictability, \cite{MarwanPerc2020} consider the white noise process as a reference process, because it has maximum entropy with no predictive information and it is also characterized by a scattered energy distribution across all scales. Generally, the larger the entropy, the more random and complex a system, and vice versa. Therefore, the white noise process, that is completely random and unpredictable, assumes the highest entropy. In \cite{Sang2015,MarwanPerc2020}, the white noise time series is supposed to be a time series with the same length of the original time series, $X$, having its same mean and standard deviation one.

Let $WE_x$ be the Wavelet Entropy of time series $X$ and let $WE_{wn}$ be the Wavelet Entropy of white noise. Then \cite{MarwanPerc2020} proposed the Wavelet Energy Entropy Measure (WEEM) as follows:
\begin{equation}
\label{eq:weem}
WEEM=1-e^{WE_x-WE_{wn}}\, .
\end{equation}
We observe that:
\begin{itemize}
    \item The value of WEEM ranges from $0$ to $1$.
    \item If WEEM is close to one, the entire energy of $X$ is concentrated around few scales and then the time series has high intrinsic predictability (because $WE_{wn}\gg WE_x$).
    \item If WEEM is close to zero, the entire energy of $X$ is scattered across all scales (similar to that of a white noise process) and then $X$ has a very low intrinsic predictability (because $WE_{wn}\approx WE_x$).
\end{itemize}
The pseudo-code of WEEM is described by Algorithm \ref{alg:weem} in the Appendix, where we adopted a slightly different approach to the definition of WEEM than that of \cite{MarwanPerc2020}. \cite{MarwanPerc2020} need a time series representing white noise, in our paper instead we need only its entropy. Moreover, since the entropy is maximum if all outcomes are equally probable (scattered on all scales, i.e. uncertainty is maximum when all possible events are equiprobable), this means that WE in Eq. (\ref{eq:WE}) is always less than or equal to $\ln J$. 
Hence, for our purposes it is sufficient to set $WE_{wn}=\ln J$.

 \subsection{Cross wavelet energy entropy measure (CWEEM)}
\label{CWEEM}
Moreover, given two time series $X$ and $Y$, we move one step forward with the aim of quantifying the intrinsic predictability of $X$ given $Y$ and viceversa. Let us now suppose that we have two different probability distributions $\{E_j^{(x)}\}$ and $\{E_j^{(y)}\}$, such that $\sum_{j=1}^J E_j^{(x)}=\sum_{j=1}^J E_j^{(y)}=1$. In this case, they represent by scales the probability distribution of the wavelet energy for the two different time series $X$ and $Y$. We use the Kullback-Leibler entropy, as defined by \cite{KullbackLeibler}, as:
\begin{equation}
\label{eq:KL}
WE_{y|x}=\sum_{j=1}^J E_j^{(y)}\, \ln\left(\frac{E_j^{(y)}}{E_j^{(x)}}\right)\, ,
\end{equation}
where $WE_{y|x}$ is a wavelet entropy which gives a measure of the degree of similarity of the distribution $\{E_j^{(y)}\}$ with respect to the distribution $\{E_j^{(x)}\}$, taken as a reference distribution. Note that $WE_{y|x}$ is positive and vanishes only if $E_j^{(y)}=E_j^{(x)}$.

Then, generalizing the WEEM proposed bx \cite{MarwanPerc2020}, we introduce the \emph{Cross Wavelet Energy Entropy Measure} (CWEEM) as follows:
\begin{equation}
\label{eq:cweem}
CWEEM=1-2^{WE_{y|x}-WE_{wn}}\, .
\end{equation}
We observe that:
\begin{itemize}
    \item The value of CWEEM still ranges from $0$ to $1$.
    \item It tells us whether the knowledge of $X$ reduces uncertainty about $Y$. In the strongest form of this argument, given two sources (series), if $Y$ is a deterministic function of $X$, then knowing $X$ lets one know the value of $Y$. Then $Y$ will have a high intrinsic predictability given $X$, $WE_{wn}\gg WE_{y|x}$ and CWEEM will be near to one.
    \item If, given $X$, we can't predict $Y$, then this will have a very low intrinsic predictability. In other words, if $Y$ is independent from $X$, then the knowledge of the latter does not give any information about the former. Hence $WE_{wn}\approx WE_{y|x}$ and CWEEM is near to 0.
\end{itemize}
The pseudo-code of CWEEM is described by Algorithm \ref{alg:cweem}.




\section{Results and discussion}
\label{results}
\subsection{Wavelet coherence and wavelet phase difference}
\label{rescoh}
In this section, we show the results obtained by employing the wavelet based methods introduced so far.
 As a first step, the oil-food price relationship will be assessed first with wavelet coherence. As a second step, the wavelet coherence will be made for food and each of the variables discussed in Sec. \ref{data}, such as the demand from emerging economies, the presence of commodity indexes, the financial stress and the global economic activity. Moreover, we will apply the methods to the case of oil. Consequently, it will be observed  how these variables are locally correlated to food and oil. 
As a final step, the partial wavelet coherence will be applied to see if the relationship between food and oil can be attributed to the above mentioned variables. 

In the wavelet coherence graphs, the white curve denotes the cone of influence (CoI). Indeed, the CWT has border distortions when applied to finite length time series as $s$ increases, thus the CoI is introduced because errors occur at the beginning and at the end of the wavelet power spectrum \citep{doi:10.1175/1520-0477(1998)079<0061:APGTWA>2.0.CO;2,aguiar2011oil}. 
The CoI is a region in which the edge effects can not be ignored and thus the interpretation of the results has to be done carefully \citep{grinsted:hal-00302394,aguiar2011oil}.  

\begin{figure}[hbt]
    \subfloat[Wavelet coherence food price index-GSCI]{\includegraphics[scale=0.25]{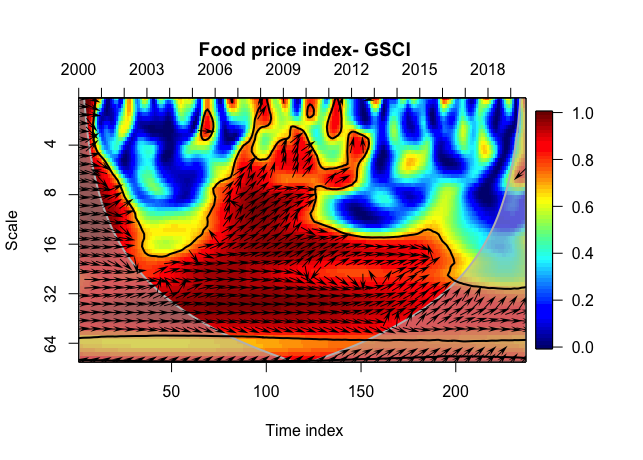}\label{fig:A}}\hspace{2mm}
    \subfloat[Wavelet coherence oil price index-GSCI]{\includegraphics[scale=0.25]{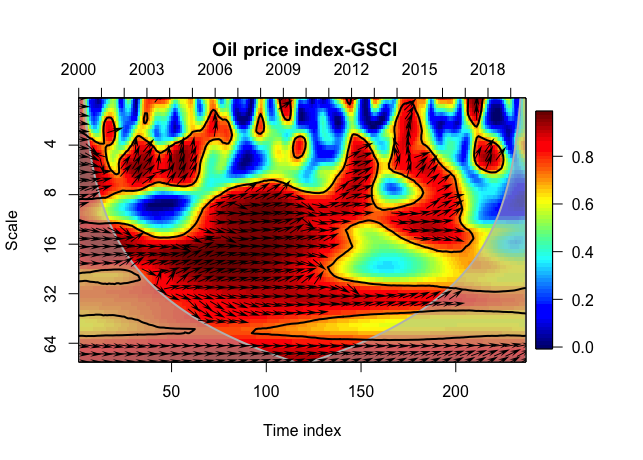}\label{fig:B}}\\
    \subfloat[Wavelet coherence oil price index-food price index ]{\includegraphics[scale=0.25]{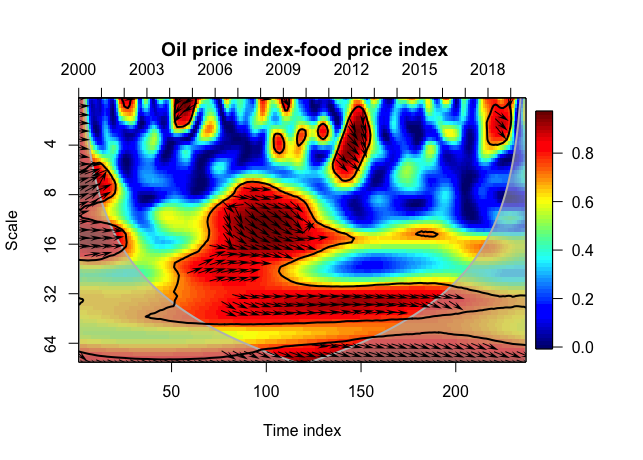}\label{fig:C}}\hspace{2mm}
    \subfloat[Partial wavelet coherence oil price index-  food price index controlling for the effect of GSCI]{\includegraphics[scale=0.25]{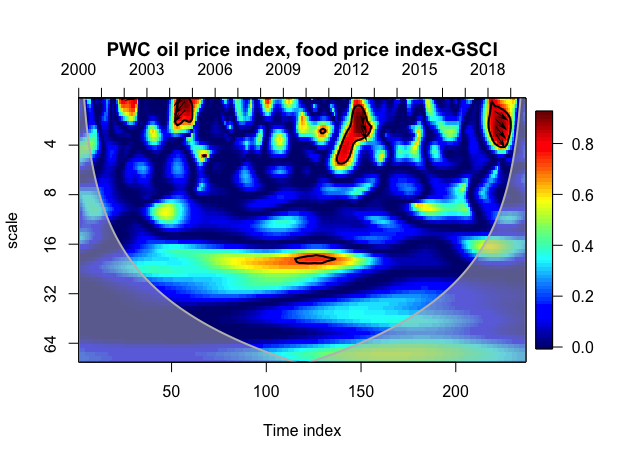}\label{fig:D}}
    \caption{Wavelet coherence between oil price index and food price index and partial wavelet coherence between  oil price index and food price index by partialling out the effect of GSCI total return index. The vertical axis refers to the  scales}
    \label{fig:resultsgsci}
\end{figure}

\begin{figure}[hbt]
    \subfloat[Wavelet coherence oil price index- Ted spread]{\includegraphics[scale=0.25]{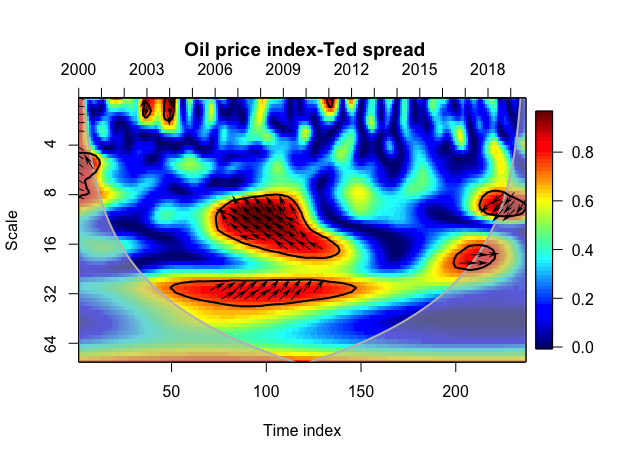}\label{fig:E2}}\hspace{2mm}
    \subfloat[Wavelet coherence food price index-Ted spread]{\includegraphics[scale=0.25]{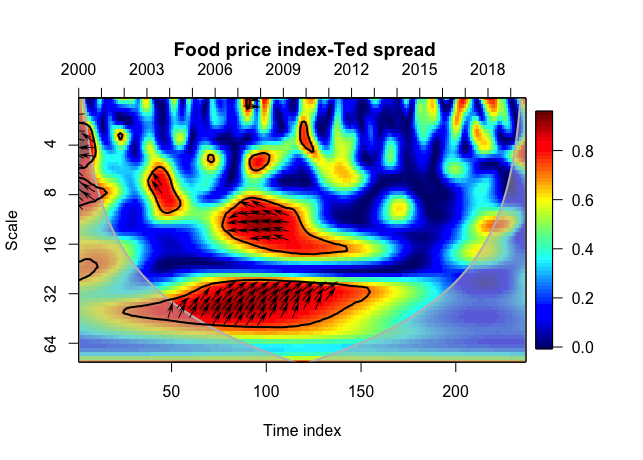}\label{fig:F2}}\\
    \subfloat[Partial wavelet coherence oil price index-food price index controlling for the effect of Ted spread  ]{\includegraphics[scale=0.25]{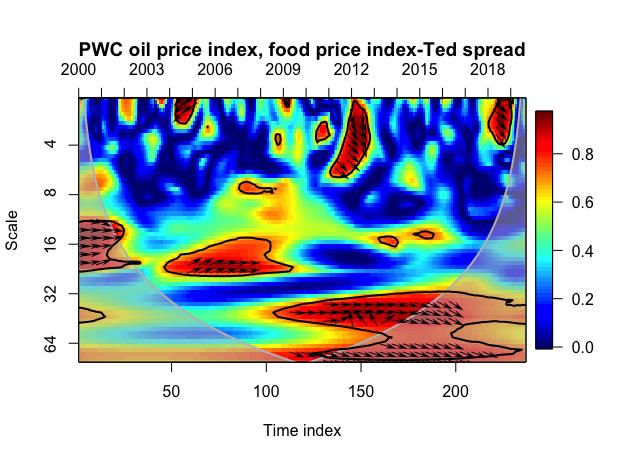}\label{fig:G2}}\hspace{2mm}
    \subfloat[Wavelet coherence food price index-MSCI index]{\includegraphics[scale=0.25]{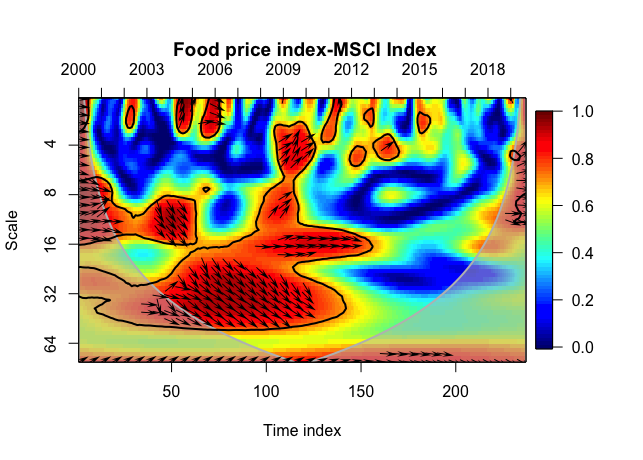}\label{fig:H2}}\hspace{2mm}
    \caption{Wavelet coherence between oil price index and TED spread and between food price index and TED spread and between food price index and MSCI emerging market index. The vertical axis refers to the scale}
    \label{fig:TED}
\end{figure}

\begin{figure}[hbt]
   \subfloat[Wavelet coherence oil price index- MSCI index]{ \includegraphics[scale=0.25]{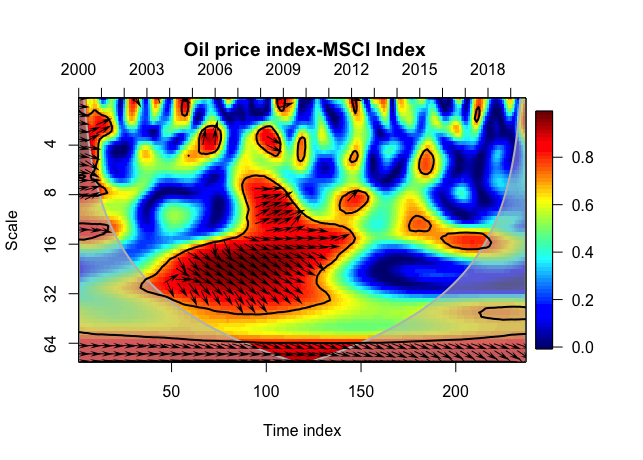}\label{fig:I}}\hspace{2mm}
   \subfloat[Partial wavelet coherence oil price index-food price index controlling for the effect of MSCI index]{\includegraphics[scale=0.25]{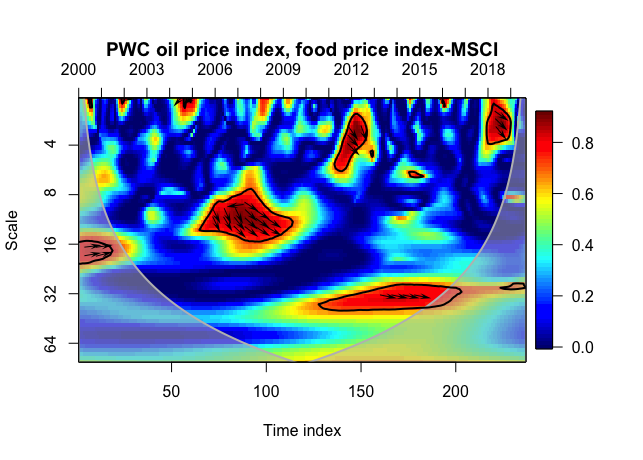}\label{fig:L}}\\
   \subfloat[Wavelet coherence food price index-Kilian economic index]{\includegraphics[scale=0.25]{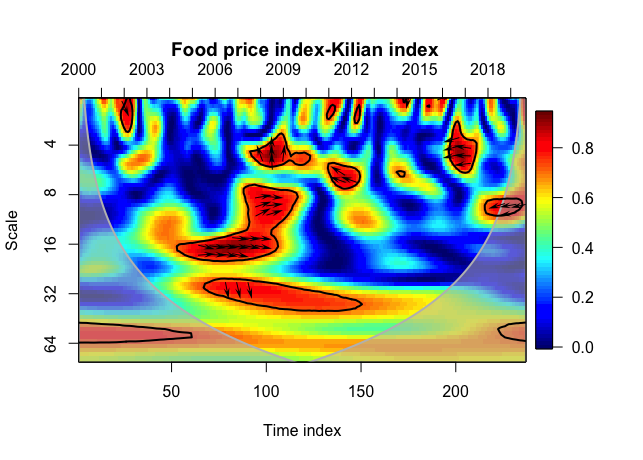}\label{fig:M}}\hspace{2mm}
   \subfloat[Wavelet coherence oil price index-Kilian economic index]{ \includegraphics[scale=0.25]{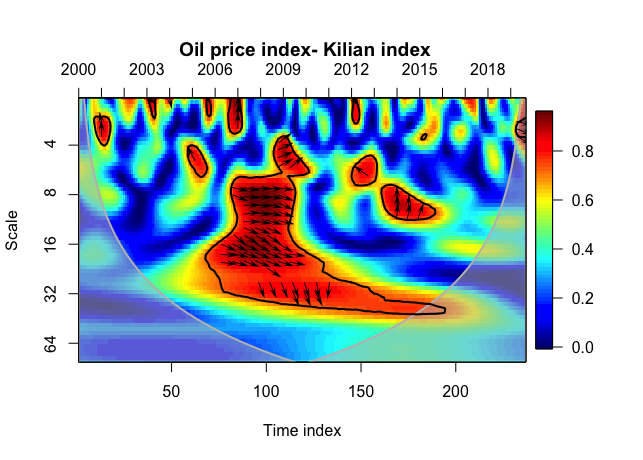}\label{fig:N}}\hspace{2mm}
   \caption{Wavelet coherence between the oil price index and MSCI emerging market index and between food price index and Kilian economic index and partial wavelet coherence between between the oil price index, food price index by partialling out the effect of MSCI emerging market index. The vertical axis refers to the  scales}
    \label{fig:resultsmsci}
\end{figure}

\begin{figure}[hbt]
    \centering
    \includegraphics[scale=0.4]{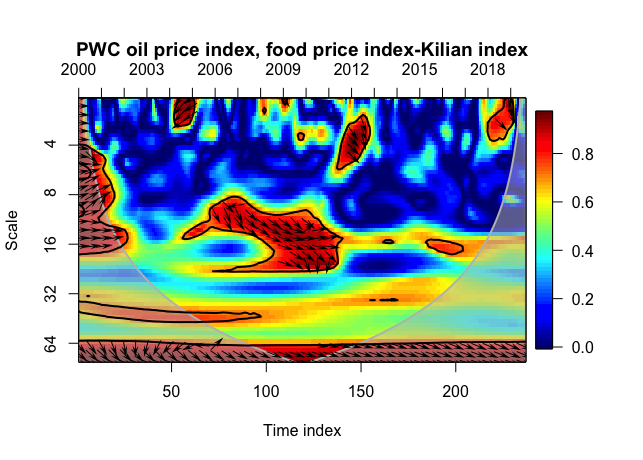}
    \caption{Wavelet coherence between oil price index and Kilian economic index. The vertical axis refers to the  scales}
    \label{fig:resultskilian}
\end{figure}

In all the Figures, the tick black contours represent significance at the level of 5\% and the significance values are determined using Monte Carlo Simulation. 
The color bar in the wavelet coherence graphs  represents the strength of co-movements between two time series at each scale across time. The red color denotes that a strong linear relationship is found, whereas the blue color denotes a weak linear relationship.

The interpretation of the results have to be done recalling that short wavelet scales correspond to co-movements which are strong at high frequencies (i.e. in the short run), whereas long wavelet scales correspond to co-movements which are strong at low frequency (i.e. in the long run). Furthermore, the series co-move positively (negatively) when they are in-phase (out of phase), as described in Sec. \ref{phase}. 
We must recall that low frequency movements (high frequency movements) correspond to global information (transient information) on the time series dynamics.
 
Fig. \ref{fig:resultsgsci} shows  results of the wavelet coherence and partial wavelet coherence for the oil price, the food price index and the GSCI total return index. 

The wavelet coherence between the food price index and the GSCI total return index in Fig. \ref{fig:resultsgsci} (a) indicates a  strong in-phase relationship  across different scales. More in details, the two series  co-move positively at long scales (16-64 periods) with no clear lead/lag relationship during the period 2003-2015. At short scales, the local correlation between the two series almost disappears. 

In Fig. \ref{fig:resultsgsci} (b), similar results are found for the wavelet coherence between the oil price index and the GSCI total return index. 

Fig. \ref{fig:resultsgsci} (c) represents the results of the wavelet coherence between  the food price index and the oil price index. In this case,  the two series co-move positively at short scales (8-16) with the food price index leading the oil price index during the period 2006-2009. At long scales (32-64), the two series still co-move positively, but no clear lead/lag relationship is found.


The evidence that the food price index leads the oil price index in the period 2006-2009 seems counter-intuitive. However, the development of biofuel industry together with the food crisis could be an explanation for this result.

In Fig. \ref{fig:resultsgsci} (d), we employ  the partial wavelet coherence between the oil price index and the food price index controlling for the effect of the GSCI total return index. In this case, it is observed that co-movements between the oil price index and the food price index significantly decrease. Thus, we have found evidence that the presence of commodity index is one of the responsible for the significant co-movement between oil and food.

Fig. \ref{fig:TED} (a) shows the wavelet coherence between the oil price index and the TED spread. These two series are out of phase at short wavelet scales (8-16) with the TED spread leading the oil price index during the period 2006-2008. This result is consistent with the fact that oil prices steadily declined in the second half of 2008 because of the financial crisis, therefore when the TED spread was increasing \citep{www5}. This result explains that financial distress was able to lead crude oil prices in the short run, therefore it has a short-term impact.  At long scales (32-64), the two series are in-phase (positive co-movement) with the oil price index leading the TED spread.  In Fig. \ref{fig:TED} (b),  similar results are found for the wavelet coherence  between the food price index and the TED spread. However, the food price index and the TED spread are out of phase (negative co-movement) at short wavelet scales (8-16) with no lead/lag relationship during the period 2006-2008. At long wavelet scales (32-64), the two series are in-phase (positive co-movement) with the food price index leading the TED spread. 

Fig. \ref{fig:TED} (c) shows  the partial wavelet coherence between the oil price index and the food price index that control the effect of the TED spread. In this case, co-movement between food and oil decreases, but less than when we control for the effect of GSCI total return index, as shown in Fig. \ref{fig:resultsgsci} (d). As a result, the systemic financial distress represented by the TED spread explained a small portion of the co-movement between food and oil that is  mainly related to the period of financial turmoil.

Fig. \ref{fig:resultsmsci} (a) represents the results of the wavelet coherence between the food price index and the MSCI Emerging Market Index. More in details,  the two series are in-phase (positive co-movement) at long wavelet scales (32-64) with MSCI Emerging Market Index leading the food price index during the period 2003-2006. 

Fig. \ref{fig:resultsmsci} (b) shows similar results for the wavelet coherence between the oil price and the MSCI emerging market index. Indeed, the two series are in-phase (positive co-movement) at long wavelet scales (16-64) with the MSCI emerging market index leading the oil price index during the period 2003-2009. 


Fig. \ref{fig:resultsmsci} (c) represents the partial wavelet coherence between the oil price index and the food price index controlling for the effect of MSCI index. In this case,  we notice that co-movement decreases. This result suggests that the demand from emerging economies may have induced joint price movement between food and oil. However, compared to \ref{fig:resultsgsci} (d), it is observed that co-movement decreases less than when we control for the effect of GSCI total return index. Consequently, this suggests that GSCI  explains more about the oil-food price relationship than the demand from emerging economies.

In Fig. \ref{fig:resultskilian},  the results of the wavelet coherence for food and oil using the Kilian economic index are shown.  Fig.  \ref{fig:resultskilian} (a) and (b) show that co-movement is greater for the oil price index and the Kilian economic index, rather than when considering the food price index. More in details, Fig.  \ref{fig:resultskilian} (a) and (b) show in-phase relationship (positive co-movement) at short (8-16) and long wavelet scales (16-32) with no clear lead/lag relationship during the period 2006-2009. 


Fig.  \ref{fig:resultskilian} shows the partial wavelet coherence between oil and food controlling for the effect of Kilian economic index. This result shows that co-movement is still present.


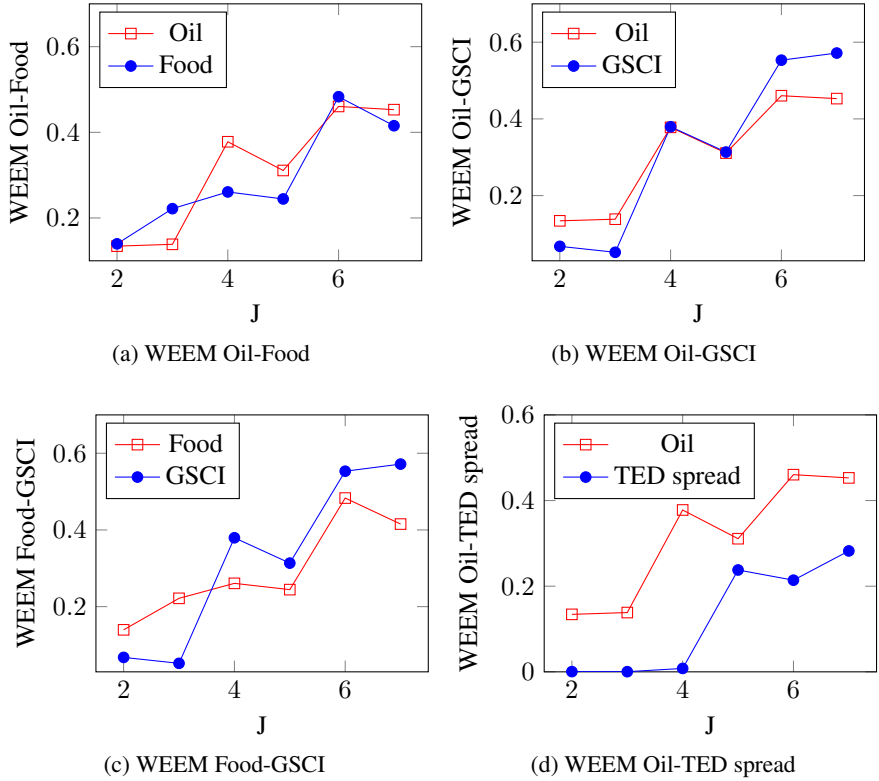
\begin{figure}[hbt]
    \subfloat[WEEM Oil-Food]{\begin{tikzpicture}
	\begin{axis}[
			xlabel=J,
	ylabel=WEEM Oil-Food,ymin=0.10,ymax=0.70,legend pos=north west,width=6cm, height=5cm]
\addplot[color=red,mark=square] coordinates {
	(2,0.1342235)
	(3,0.1384597)
	(4,0.3779662)
	(5,0.3108987)
	(6,0.4604113)
	(7,0.4529047)
	};
	
\addplot[color=blue,mark=*] 
coordinates {
	(2,0.139547)
	(3,0.2217195)
	(4,0.2607548)
	(5,0.2443900)
	(6,0.4831498)
	(7,0.4153400)

};
\legend{Oil,Food}
\label{WEEMoilfood}
	\end{axis}
\end{tikzpicture}\label{fig:Aweem}}\hspace{2mm}
    \subfloat[WEEM Oil-GSCI]{\begin{tikzpicture}
\begin{axis}[
	xlabel=J,
	ylabel=WEEM Oil-GSCI,ymin=0.03,ymax=0.70,legend pos=north west,width=6cm, height=5cm]
\addplot[color=red,mark=square] coordinates {
    (2,0.1342235)
	(3,0.1384597)
	(4,0.3779662)
	(5,0.3108987)
	(6,0.4604113)
	(7,0.4529047)

};

\addplot[color=blue,mark=*] 
coordinates {
	(2,0.06768948)
	(3,0.05217124)
	(4,0.37956328)
	(5,0.3135298)
	(6,0.55316563)
	(7,0.57170886)

};

\legend{Oil,GSCI}
\label{WEEMoilGSCI}
\end{axis}
\end{tikzpicture}\label{fig:Bweem}}\\
    \subfloat[WEEM Food-GSCI ]{
\begin{tikzpicture}
\begin{axis}[
	xlabel=J,
	ylabel=WEEM Food-GSCI,ymin=0.03,ymax=0.70,legend pos=north west,width=6cm, height=5cm]
\addplot[color=red,mark=square] coordinates {
  	(2,0.139547)
	(3,0.2217195)
	(4,0.2607548)
	(5,0.2443900)
	(6,0.4831498)
	(7,0.4153400)  

};

\addplot[color=blue,mark=*] 
coordinates {
	(2,0.06768948)
	(3,0.05217124)
	(4,0.37956328)
	(5,0.3135298)
	(6,0.55316563)
	(7,0.57170886)

};

\legend{Food,GSCI}
\label{WEEMoilGSCI2}
\end{axis}
\end{tikzpicture}\label{fig:Cweem}}\hspace{2mm}
    \subfloat[WEEM Oil-TED spread]{\begin{tikzpicture}
\begin{axis}[
	xlabel=J,
	ylabel=WEEM Oil-TED spread,ymin=0,ymax=0.60,legend pos=north west,width=6cm, height=5cm]
\addplot[color=red,mark=square] coordinates {
    (2,0.1342235)
	(3,0.1384597)
	(4,0.3779662)
	(5,0.3108987)
	(6,0.4604113)
	(7,0.4529047)

};

\addplot[color=blue,mark=*] 
coordinates {
	(2,0.0005174642)
	(3,0.0003494771)
	(4,0.0079581662 )
	(5,0.2378619460)
	(6,0.21404299927)
	(7,0.2823302778)

};
\legend{Oil,TED spread}
\label{WEEMoilTedspread}
\end{axis}
\end{tikzpicture}\label{fig:Dweem}}
    \caption{Comparison of the Wavelet Energy Entropy Measure of oil price index, food price index, GSCI total return index and TED spread}
    \label{fig:resultsWE1}
\end{figure}

In conclusion, the analysis performed so far  supports the argument that  the demand from emerging markets and the GSCI total return index are leading causes of the oil-food price co-movement.


\subsection{Wavelet entropy: WEEM and CWEEM}
\label{reswe}

In this section, we discuss the results for the WEEM estimates. Our aim is to estimate the predictability of the time series at different scales. We wonder if  the variables that mostly co-move with both food and oil may have a similar intrinsic predictability structure to that of oil and food.

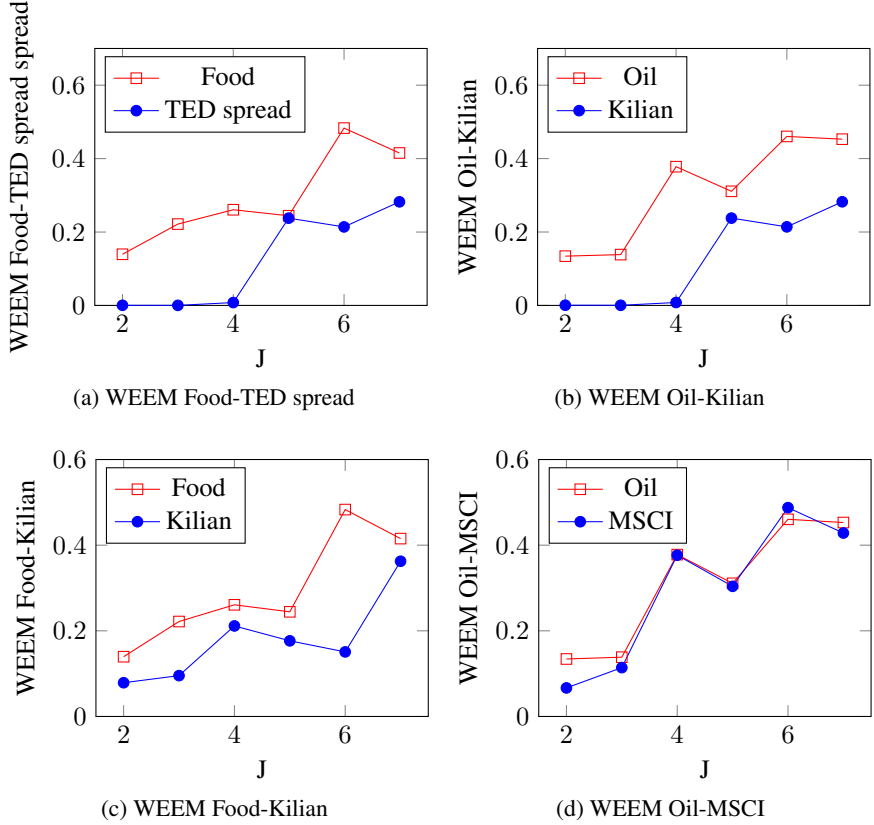
\begin{figure}[hbt]
    \subfloat[WEEM Food-TED spread]{\begin{tikzpicture}
	\begin{axis}[
			xlabel=J,
	ylabel=WEEM Food-TED spread spread,ymin=0.0002,ymax=0.70,legend pos=north west,width=6cm, height=5cm]
\addplot[color=red,mark=square] coordinates {
	(2,0.139547)
	(3,0.2217195)
	(4,0.2607548)
	(5,0.2443900)
	(6,0.4831498)
	(7,0.4153400)  

};
\addplot[color=blue,mark=*] 
coordinates {
(2,0.0005174642)
	(3,0.0003494771)
	(4,0.0079581662 )
	(5,0.2378619460)
	(6,0.21404299927)
	(7,0.2823302778)	

};
\legend{Food,TED spread}
\label{WEEMfoodTedspread}
	\end{axis}
\end{tikzpicture}\label{fig:E}}\hspace{2mm}
    \subfloat[WEEM Oil-Kilian]{\begin{tikzpicture}
\begin{axis}[
	xlabel=J,
	ylabel= WEEM Oil-Kilian,ymin=0,ymax=0.70,legend pos=north west,width=6cm, height=5cm]
\addplot[color=red,mark=square] coordinates {
    (2,0.1342235)
	(3,0.1384597)
	(4,0.3779662)
	(5,0.3108987)
	(6,0.4604113)
	(7,0.4529047)

};

\addplot[color=blue,mark=*] 
coordinates {
	(2,0.0005174642)
	(3,0.0003494771)
	(4,0.0079581662 )
	(5,0.2378619460)
	(6,0.21404299927)
	(7,0.2823302778)	

};

\legend{Oil,Kilian}
\label{WEEMoilKilian}
\end{axis}
\end{tikzpicture}\label{fig:F}}\\
    \subfloat[WEEM Food-Kilian]{
\begin{tikzpicture}
\begin{axis}[
	xlabel=J,
	ylabel=WEEM Food-Kilian,ymin=0,ymax=0.60,legend pos=north west,width=6cm, height=5cm]
\addplot[color=red,mark=square] coordinates {
    (2,0.139547)
    (3,0.2217195)
	(4,0.2607548)
	(5,0.2443900)
	(6,0.4831498)
	(7,0.4153400)   

};

\addplot[color=blue,mark=*] 
coordinates {
	(2,0.0787904)
	(3,0.09535656)
	(4,0.211417367)
	(5,0.17662872)
	(6,0.15075750)
	(7,0.36247102)

};

\legend{Food,Kilian}
\label{WEEMFoodKilian}
\end{axis}
\end{tikzpicture}\label{fig:G}}\hspace{2mm}
    \subfloat[WEEM Oil-MSCI]{\begin{tikzpicture}
\begin{axis}[
	xlabel=J,
	ylabel=WEEM Oil-MSCI,ymin=0,ymax=0.60,legend pos=north west,width=6cm, height=5cm]
\addplot[color=red,mark=square] coordinates {
    (2,0.1342235)
	(3,0.1384597)
	(4,0.3779662)
	(5,0.3108987)
	(6,0.4604113)
	(7,0.4529047)

};

\addplot[color=blue,mark=*] 
coordinates {
    (2,0.06664162)
	(3,0.11406969)
	(4,0.37661903)
	(5,0.30390290)
	(6,0.48762835)
	(7,0.42834170)

};
\legend{Oil,MSCI}
\label{WEEMoilMSCI}
\end{axis}
\end{tikzpicture}\label{fig:H}}
    \caption{Comparison of the Wavelet Energy Entropy Measure of Food price index, TED spread, Kilian economic index and MSCI emerging market index}
    \label{fig:resultsWE2}
\end{figure}


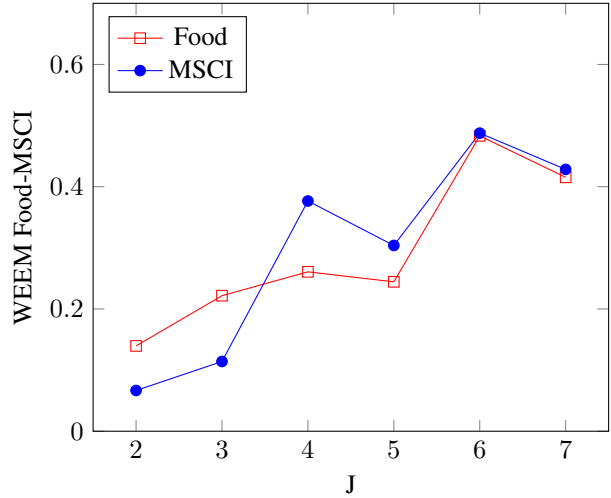
\begin{figure}[hbt]
\centering
\begin{tikzpicture}
\begin{axis}[
	xlabel=J,
	ylabel=WEEM Food-MSCI,ymin=0,ymax=0.70,legend pos=north west]
\addplot[color=red,mark=square] coordinates {
    (2,0.139547)
    (3,0.2217195)
	(4,0.2607548)
	(5,0.2443900)
	(6,0.4831498)
	(7,0.4153400)   
};
\addplot[color=blue,mark=*] 
coordinates {
	(2,0.06664162)
	(3,0.11406969)
	(4,0.37661903)
	(5,0.30390290)
	(6,0.48762835)
	(7,0.42834170)
};
\legend{Food, MSCI}
\end{axis}
\end{tikzpicture}
\caption{Comparison of Wavelet Entropy Energy Measure of Food price index and MSCI emerging market index}
\label{WEEMfoodMSCI}
\end{figure}

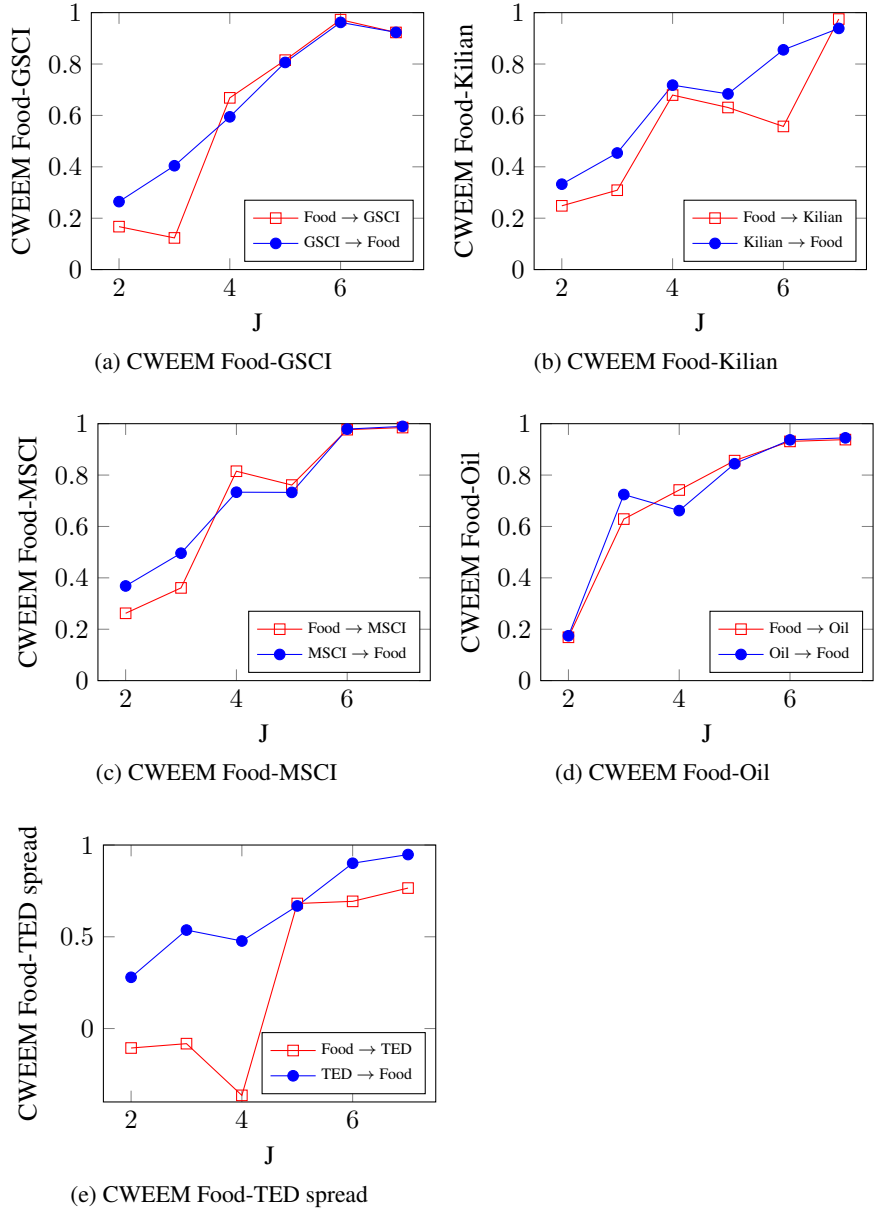
\begin{figure}[hbt]
    \subfloat[CWEEM Food-GSCI]{\begin{tikzpicture}
	\begin{axis}[
			xlabel=J,
	ylabel=CWEEM Food-GSCI,ymin=0.0002,ymax=1,legend pos=south east,width=6cm, height=5cm]
\addplot[color=red,mark=square] coordinates {
	  (2,0.167239774961846)
(3,0.123331249138754)
(4,0.668392069131601)
(5,0.815043005313239)
(6,0.972509392483318)
(7,0.922871522292503)
};
\addplot[color=blue,mark=*] 
coordinates {
(2,0.264538091013907)
(3,0.404207519779306)
(4,0.594475049031651)
(5,0.806162557080938)
(6,0.96197036548638)
(7,0.92323279774029)
};
\legend{\tiny Food $\rightarrow$ GSCI, \tiny GSCI $\rightarrow$ Food}
\label{CWEEMfoodGSCI}
	\end{axis}
\end{tikzpicture}}\hspace{2mm}
    \subfloat[CWEEM Food-Kilian]{\begin{tikzpicture}
\begin{axis}[
	xlabel=J,
	ylabel= CWEEM Food-Kilian,ymin=0,ymax=1,legend pos=south east,width=6cm, height=5cm]
\addplot[color=red,mark=square] coordinates {
    (2,0.248136110810784)
(3,0.308863501076934)
(4,0.679130940348144)
(5,0.630820591912761)
(6,0.557110889627622)
(7,0.975033277279189)
};

\addplot[color=blue,mark=*] 
coordinates {
	(2,0.332347747022052)
(3,0.453706346674054)
(4,0.717839474820813)
(5,0.683702768860339)
(6,0.855265769606204)
(7,0.938666877813068)
};
\legend{\tiny Food $\rightarrow$ Kilian, \tiny Kilian $\rightarrow$ Food}
\label{CWEEMFoodKilian}
\end{axis}
\end{tikzpicture}}\\
    \subfloat[CWEEM Food-MSCI]{
\begin{tikzpicture}
\begin{axis}[
	xlabel=J,
	ylabel=CWEEM Food-MSCI,ymin=0,ymax=1,legend pos=south east,width=6cm, height=5cm]
\addplot[color=red,mark=square] coordinates {
(2,0.262088681227271)
(3,0.360846568974678)
(4,0.814952171759124)
(5,0.761052420719265)
(6,0.976714372956618)
(7,0.984896834160876)
};

\addplot[color=blue,mark=*] 
coordinates {
(2,0.368389952929276)
(3,0.49597911030639)
(4,0.733317910165484)
(5,0.732619453271213)
(6,0.978983204718182)
(7,0.989495183478325)
};
\legend{\tiny Food $\rightarrow$ MSCI, \tiny MSCI $\rightarrow$ Food}
\label{CWEEMFoodMSCI}
\end{axis}
\end{tikzpicture}}\hspace{2mm}
    \subfloat[CWEEM Food-Oil]{\begin{tikzpicture}
\begin{axis}[
	xlabel=J,
	ylabel=CWEEM Food-Oil,ymin=0,ymax=1,legend pos=south east,width=6cm, height=5cm]
\addplot[color=red,mark=square] coordinates {
(2,0.168651130515019)
(3,0.62872603729406)
(4,0.741815269476863)
(5,0.856168937971802)
(6,0.931042814703815)
(7,0.937994209008454)
};

\addplot[color=blue,mark=*] 
coordinates {
(2,0.174552474098708)
(3,0.724199744815062)
(4,0.661706746431526)
(5,0.844134812986943)
(6,0.937000373259856)
(7,0.944869008203149)
};
\legend{\tiny Food $\rightarrow$ Oil, \tiny Oil $\rightarrow$ Food}
\label{CWEEMoilFood}
\end{axis}
\end{tikzpicture}}\\
    \subfloat[CWEEM Food-TED spread]{
\begin{tikzpicture}
\begin{axis}[
	xlabel=J,
	ylabel=CWEEM Food-TED spread,ymin=-0.4,ymax=1,legend pos=south east,width=6cm, height=5cm]
\addplot[color=red,mark=square] coordinates {
(2,-0.106032948750623)
(3,-0.0823228206993536)
(4,-0.364015535016309)
(5,0.681420234403336)
(6,0.693107561187474)
(7,0.765524014431022)
};

\addplot[color=blue,mark=*] 
coordinates {
(2,0.279290819831126)
(3,0.536652259823127)
(4,0.476977678735976)
(5,0.668115050008687)
(6,0.900779958161271)
(7,0.948069310251461)
};
\legend{\tiny Food $\rightarrow$ TED, \tiny TED $\rightarrow$ Food}
\label{CWEEMFoodTED}
\end{axis}
\end{tikzpicture}}
    \caption{Comparison of the Cross Wavelet Energy Entropy Measure of Food price index vs GSCI, Oil price index, TED spread, Kilian economic index and MSCI emerging market index}
    \label{fig:resultsCWEEM1}
\end{figure}

\begin{figure}[hbt]
    \subfloat[CWEEM Oil-GSCI]{\begin{tikzpicture}
	\begin{axis}[
			xlabel=J,
	ylabel=CWEEM Oil-GSCI,ymin=0,ymax=1.1,legend pos=south east,width=6cm, height=5cm]
\addplot[color=red,mark=square] coordinates {
(2,0.107020980359538)
(3,0.310203329165964)
(4,0.746030919569996)
(5,0.773535888121308)
(6,0.993715969963598)
(7,0.868550027097617)
};
\addplot[color=blue,mark=*] 
coordinates {
(2,0.188160199791752)
(3,0.464173789499915)
(4,0.753979346167095)
(5,0.775612545585157)
(6,0.987705492145158)
(7,0.865579038289929)
};
\legend{\tiny Oil $\rightarrow$ GSCI, \tiny GSCI $\rightarrow$ Oil}
\label{CWEEMOilGSCI}
	\end{axis}
\end{tikzpicture}}\hspace{2mm}
    \subfloat[CWEEM Oil-Kilian]{\begin{tikzpicture}
\begin{axis}[
	xlabel=J,
	ylabel= CWEEM Oil-Kilian,ymin=0,ymax=1,legend pos=south east,width=6cm, height=5cm]
\addplot[color=red,mark=square] coordinates {
(2,0.223843976846199)
(3,0.387522818527135)
(4,0.564258081034312)
(5,0.772488468636552)
(6,0.725898925327484)
(7,0.954872487382359)
};

\addplot[color=blue,mark=*] 
coordinates {
(2,0.299536302948182)
(3,0.443690014170427)
(4,0.712830348079963)
(5,0.855391124032558)
(6,0.935099012500199)
(7,0.814165409842581)
};
\legend{\tiny Oil $\rightarrow$ Kilian, \tiny Kilian $\rightarrow$ Oil}
\label{CWEEMOilKilian}
\end{axis}
\end{tikzpicture}}\\
    \subfloat[CWEEM Oil-MSCI]{
\begin{tikzpicture}
\begin{axis}[
	xlabel=J,
	ylabel=CWEEM Oil-MSCI,ymin=0,ymax=1,legend pos=south east,width=6cm, height=5cm]
\addplot[color=red,mark=square] coordinates {
(2,0.263290912794464)
(3,0.338674484985151)
(4,0.734190567475111)
(5,0.749215847132291)
(6,0.924286398404043)
(7,0.919697148737992)
};

\addplot[color=blue,mark=*] 
coordinates {
(2,0.362167854858943)
(3,0.373473332239255)
(4,0.735494707906084)
(5,0.785345661512338)
(6,0.921209106977241)
(7,0.900984289368826)
};
\legend{\tiny Oil $\rightarrow$ MSCI, \tiny MSCI $\rightarrow$ Oil}
\label{CWEEMOilMSCI}
\end{axis}
\end{tikzpicture}}\hspace{2mm}
    \subfloat[CWEEM Oil-TED spread]{
\begin{tikzpicture}
\begin{axis}[
	xlabel=J,
	ylabel=CWEEM Oil-TED spread,ymin=-0.7,ymax=1,legend pos=south east,width=6cm, height=5cm]
\addplot[color=red,mark=square] coordinates {
(2,-0.157604805475365)
(3,-0.0223776028553588)
(4,-0.65947733651022)
(5,0.568843660272945)
(6,0.82965022306809)
(7,0.626568025976999)
};

\addplot[color=blue,mark=*] 
coordinates {
(2,0.161917281665439)
(3,0.391114805051984)
(4,0.508439064185571)
(5,0.702166161439806)
(6,0.960063189869106)
(7,0.879423705253486)
};
\legend{\tiny Oil $\rightarrow$ TED, \tiny TED $\rightarrow$ Oil}
\label{CWEEMOilTED}
\end{axis}
\end{tikzpicture}}
    \caption{Comparison of the Cross Wavelet Energy Entropy Measure of Oil price index vs GSCI, TED spread, Kilian economic index and MSCI emerging market index}
    \label{fig:resultsCWEEM2}
\end{figure}
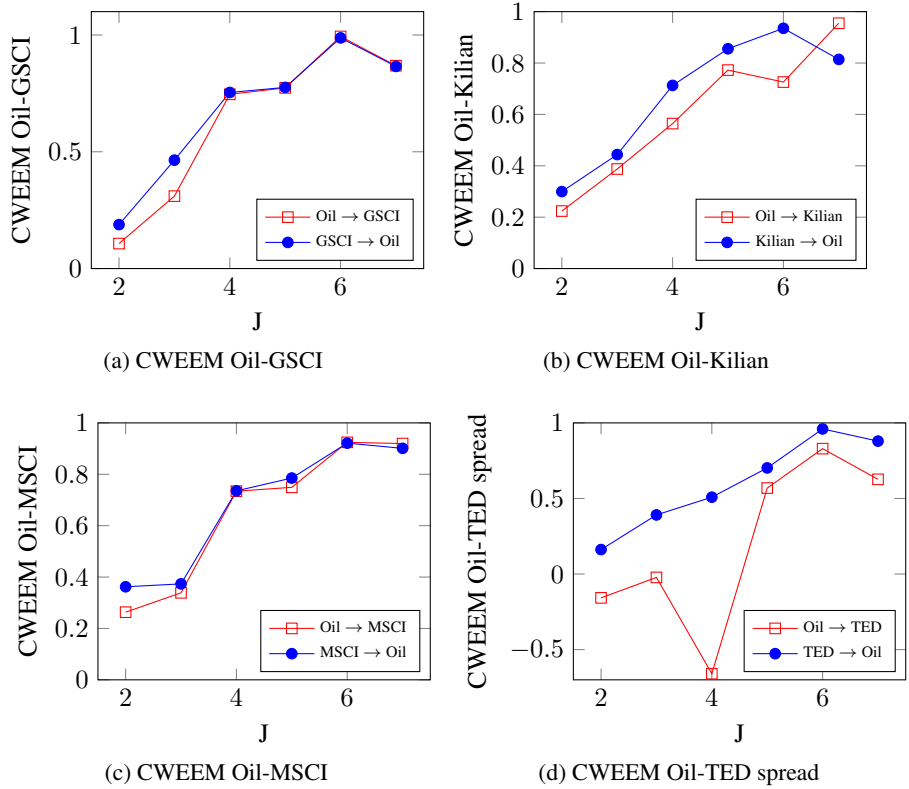

More in details, the results show how the WEEM varies as the scale $J$ varies. The scale $J$ takes integer values from $2$ to $7$. $J$ is an integer that specifies the level of the decomposition. It is such that the length of time series is at least as great as the length of the level $J$ wavelet filter, but less than the length of the level $J+1$ wavelet filter. Thus, $J \leq \log \left(\frac{N-1}{L-1}+1\right)$, where $N$ is the time series length and $L$ is an integer representing the length of the wavelet and scaling filters \citep{percival2000wavelet}.

As described in Sec. \ref{WEEM}, if WEEM is close to one, the entire energy of $X$ is concentrated around few scales and then the time series has high intrinsic predictability. On the other hand, if WEEM is close to zero, the entire energy of $X$ is scattered across all scales (similar to that of a white noise process) and then $X$ has a very low intrinsic predictability. 

We can observe from the graphs below that predictability increases as the scale increases. More in details, it increases for long wavelet scales (i.e. in the long run) and decreases for short wavelet scales (i.e. in the short run). 

Furthermore, we can note  that the behaviour of the WEEM of MSCI and GSCI is very similar to that of the WEEM of the oil price index and  the food price index as it is shown in Figs. \ref{fig:resultsWE1} (b), \ref{fig:resultsWE1} (c), \ref{fig:resultsWE2} (d) and \ref{WEEMfoodMSCI}. This evidence shows that the variables, that mostly co-move with both the food price index and  the oil price index, as shown in Sec. \ref{results}, share also their same predictability structure. 

The relation between short and long run are important from a forecasting point of view. Indeed, since the predictability increases as the scale increases, these series are more predictable in the long-run. Therefore, global information is more predictable than transient one. As a consequence, forecasting models should consider only global information of a time series for prediction.

 With respect to the CWEEM graphs, we observe in Fig. \ref{fig:resultsCWEEM1}(a) that GSCI helps to predict  the food price index at short scales, thus for high frequency movements.  Furthermore, we note in Fig. \ref{fig:resultsCWEEM1} (b) that the Kilian economic index helps to predict the dynamics of the food price index, thus global economic activity helps to predict the dynamics of the food price index for long and short scales. Fig. \ref{fig:resultsCWEEM1} (c)  shows that MSCI helps to predict the food price index for short scales, thus for high frequency movements. In Fig. \ref{fig:resultsCWEEM1} (d), it is  not clear whether food can predict oil and vice-versa since the values of the CWEEM are very similar at all scales. With respect to Fig. \ref{fig:resultsCWEEM1} (e), we must point out that the TED has a very low intrinsic predictability as can be seen in both Figs. \ref{fig:resultsWE2} (a) and \ref{fig:resultsWE1}(d), where the blue line is close to zero and has a very low value at all scales. Consequently, the results of Fig.\ref{fig:resultsCWEEM1} (e) are not reliable, as by referring to Eq. (\ref{eq:cweem}), where, in this case, we  have that   $x$  is the food price index and $y$  is the TED spread in $WE_{x|y}$. Therefore, it appears that $WE_{x|y}$ is more unpredictable than a white noise conditioned to the TED spread. In this case, it means that knowing the TED spread even increase the uncertainty about the food price index.
 Figs \ref{fig:resultsCWEEM2} (a) and (c) show the same results as Figs. \ref{fig:resultsCWEEM1} (a) and (c), thus that GSCI and MSCI help to predict  the oil price index at short scales. Furthermore, Fig. \ref{fig:resultsCWEEM2} (b) shows again that global economic activity helps to predict the dynamics of the oil price index at all scales. The problem of the low intrinsic predictability of the TED spread is also present in Fig. \ref{fig:resultsCWEEM2} (d).

\section{Conclusions and policy implications}
\label{conclusions}
This paper, using wavelet analysis, re-visits the oil-food price relationship. It assesses that the activity of commodity index investments is the main responsible for the apparent co-movement between food and oil. However, also the demand from emerging economies plays a role, but to a lower extent. Furthermore, the TED spread is only responsible for the oil-food price co-movement during the financial crisis, thus its effect cannot be considered as one of the leading causes of the oil-food price relationship. 

In addition, we find the presence for all the variables of a lead/lag relationship only at high frequency (in the short run) whereas at low frequency (in the long run) there is no clear lead/lag relationship. 

As a consequence, if there is no lead/lag relationship in the long-run, it is necessary to pay attention on transient relations: transient relations among variables, their phase-relationship along with  their lead/lag relationship could be useful for investors for diversification purposes.


The results of the WEEM show that both oil price index  and food price index share the same predictability structure  with the  S \& P GSCI and MSCI index. Furthermore, all the series are more predictable as the scale increases. Therefore, global information is more predictable than transient one, suggesting the use  of forecasting models only for global dynamics of time series.

On the other hand, the results of the CWEEM show that global economic activity is fundamental to predict the dynamics of the oil price index and the food price index, while commodity index investments and demand from emerging economies predict the oil and food price index only in the short run. 
 
As far as policy implications are concerned, some of them may be derived from our results.
The first one concerns the joint movement of prices between food and oil. 
 The oil price index and the food price co-move positively, thus an increase (decrease) in the price of oil is followed by an increase (decrease) in the price of food.  As a result, this can have a negative impact on the trade balance of those countries who rely on both oil and food imports. As suggested by \cite{HERNANDEZ2019588}, the dynamics of oil and food should prompt governments to allocate subsidy packages in energy and agricultural sector. 
 Compared to \cite{HERNANDEZ2019588}, we suggest that governments should provide subsidy packages for the commodity traded in the commodity indices to protect producers and consumers against price movements due to financial activity rather than supply or demand shortage. \textcolor{black}{Our suggestion relies on the fact} \textcolor{black}{that the in-phase relationship between the oil price index and the food price is mainly due to the activity of commodity index investments. This could help to avoid sharp} \textcolor{black}{rise or fall in these commodity prices that can, in turn, affect both the demand and supply side even if the shocks do not come from the real sector.}
 
\textcolor{black}{In a nutshell, as commodity markets has undergone a process of financialization, the government should work to limit the impact of financial activity on the commodity sector. In fact, the activity of commodity investments, that can create joint price movements even between unrelated commodities,  can increase food insecurity in the case where  prices of commodities traded in such indices may undergo sharp increase or decline.}
 
 

 \textcolor{black}{The second policy implication is about the role of commodity indices. As showed in the results and discussion section, these indices are the main responsible for the correlation between oil and food and the other variables  played a less important role. However, we must consider that all the results found are of particular interest in the short term, as the  co-movement and phase relationship are clearly identifiable at short wavelet scales. }
\textcolor{black}{This suggests that these results may be of particular relevance for policy making only when a short time horizon is considered. In fact, if we also consider the results of the CWEEM, it seems that only global economic activity can predict the dynamics of oil and food in both the short and long run, whereas GSCI is  only relevant in the short run.}






\textcolor{black}{Despite the aforementioned contributions of this study to the current literature, this paper has some limitations.}

\textcolor{black}{First of all, to address the first research question, we explored relationships among S\&P GSCI commodity index, the oil price index and the food price index. Nevertheless, S\&P GSCI index is a broad index that includes energy, metal and agriculture \textcolor{black}{commodities.} Fundamentals affecting those sub-indices are different. As future works, it could be interesting to explore the relationship of such sub-indices (like e.g., SP GSCI Softs coffee, sugar, cocoa and cotton or SP GSCI precious metal gold and silver) with oil and food price index.}

\textcolor{black}{On the second research question, we used TED spread as an indicator of financial distress. For the future works, the outcome of this paper could be compared with other series like, e.g., major stock indices, inflation and interest rate and economic policy uncertainty index.}
\textcolor{black}{In addition, for future research work it would be interesting to analyse how the COVID-19 pandemic has changed  the relationship between oil and agricultural commodities and how it has affected the  process of financialisation of commodity markets.} \textcolor{black}{In addition, the very recent and unexpected scenario of the Russia-Ukraine conflict} \textcolor{black}{may have changed the relationship between these two markets and this will be very interesting to analyse once  \textcolor{black}{sufficiently} long series of data will be available.}

\appendix
\renewcommand{\thesection}{\Alph{section}}
\section{Appendix}
\label{sec:appendix}

\subsection{Algorithms}

\begin{algorithm}
\caption{DWT computation according to pyramid algorithm, see \cite{percival2000wavelet}}
\label{alg:DTW}
\SetKwInOut{Input}{Input}
\SetKwInOut{Output}{Output}
\Input{Vector $\textbf{V}_{j-1}$ of even length $M = N/2^{j-1}$}
\Output{Vectors $\textbf{W}_{j}$ and $\textbf{V}_{j}$}
\For{$t\leftarrow 0$ \KwTo $M/2-1$}{
$u\leftarrow 2t+1$\;
$W_{j,t}\leftarrow h_0\, V_{j-1,u}$\;
$V_{j,t}\leftarrow g_0\, V_{j-1,u}$\;
\For{$n\leftarrow 1$ \KwTo $L-1$}{
$u\leftarrow u-1$\;
\If{$u<0$}{
   $u\leftarrow M-1$\;
   }
$W_{j,t}\leftarrow W_{j,t}+ h_n\, V_{j-1,u}$\;
$V_{j,t}\leftarrow V_{j,t}+ g_n\, V_{j-1,u}$\;
}
}
\end{algorithm}

\begin{algorithm}
\caption{WEEM computation}
\label{alg:weem}
\SetKwInOut{Input}{Input}
\SetKwInOut{Output}{Output}
\Input{Time series $X$, $K$ (number of Monte Carlo runs), $J$}
\Output{Wavelet Energy Entropy Measure (WEEM)}
$WE_{wn}\leftarrow \ln J$\;
$\left\{\textbf{W}_{j}'\right\}_{j=1}^J \leftarrow dwt$(X, J) \tcp{calculate wavelet coefficients of X according to Algorithm \ref{alg:DTW}}
\For{$j\leftarrow 1$ \KwTo $J$}{
$E_{j}'\leftarrow \frac{\|\textbf{W}_j'\|^2}{\sum_{j=1}^J \|\textbf{W}_j'\|^2}$\tcp{relative energy at scale j}
}
$WE_{x}\leftarrow -\sum_{j=1}^J E_j'\, \ln\left(E_j'\right)$\tcp{Calculate the wavelet entropy of $X$}
$WEEM\leftarrow 1-2^{WE_x-WE_{wn}}$\tcp{Wavelet Energy Entropy Measure}
\end{algorithm}

\begin{algorithm}
\caption{CWEEM computation}
\label{alg:cweem}
\SetKwInOut{Input}{Input}
\SetKwInOut{Output}{Output}
\Input{Time series $X$, $Y$, $K$ (number of Monte Carlo runs), $J$}
\Output{Cross Wavelet Energy Entropy Measure (CWEEM)}
$WE_{wn}=0$\;
\For{$k\leftarrow 0$ \KwTo $K$}{\tcp{Monte Carlo runs}
$WN \leftarrow$ normal distribution$\left(\operatorname{length}(X),\operatorname{mean}(X),sd=1\right)$\;
$\left\{\textbf{W}_{j}\right\}_{j=1}^J \leftarrow dwt$(WN, J) \tcp{calculate wavelet coefficients of WN according to Algorithm \ref{alg:DTW}}
\For{$j\leftarrow 1$ \KwTo $J$}{
$E_{j}\leftarrow \frac{\|\textbf{W}_j\|^2}{\sum_{j=1}^J \|\textbf{W}_j\|^2}$\tcp{relative energy at scale j}
}
$\overline{WE}_{wn}\leftarrow -\sum_{j=1}^J E_j\, \ln\left(E_j\right)$\tcp{Calculate the wavelet entropy of WN}
\If{$\overline{WE}_{wn}>WE_{wn}$}{
   $WE_{wn}\leftarrow \overline{WE}_{wn}$\tcp{white noise process assumes the highest entropy}
}
}
$\left\{\textbf{W}_{j}^{(x)}\right\}_{j=1}^J \leftarrow dwt$(X, J) \tcp{calculate wavelet coefficients of X according to Algorithm \ref{alg:DTW}}
$\left\{\textbf{W}_{j}^{(y)}\right\}_{j=1}^J \leftarrow dwt$(Y, J) \tcp{calculate wavelet coefficients of Y according to Algorithm \ref{alg:DTW}}
\For{$j\leftarrow 1$ \KwTo $J$}{
$E_{j}^{(x)}\leftarrow \frac{\|\textbf{W}_j^{(x)}\|^2}{\sum_{j=1}^J \|\textbf{W}_j^{(x)}\|^2}$\tcp{relative energy at scale j}
$E_{j}^{(y)}\leftarrow \frac{\|\textbf{W}_j^{(y)}\|^2}{\sum_{j=1}^J \|\textbf{W}_j^{(y)}\|^2}$\tcp{relative energy at scale j}
}
$WE_{y|x}\leftarrow \sum_{j=1}^J E_j^{(y)}\, \ln\left(\frac{E_j^{(y)}}{E_j^{(x)}}\right)$\tcp{Calculate the wavelet Kullback-Leibler entropy}
$CWEEM\leftarrow 1-2^{WE_{y|x}-WE_{wn}}$\tcp{Cross Wavelet Energy Entropy Measure}
\end{algorithm}


\newpage


%
%

\end{document}